\documentclass[english]{article}
\usepackage[T1]{fontenc}
\usepackage[latin9]{inputenc}
\usepackage{float}
\usepackage{amsmath}
\usepackage{graphicx}
\usepackage{setspace}
\usepackage{comment}
\usepackage{xcolor}
\usepackage[authoryear]{natbib}
\makeatletter

\floatstyle{ruled}
\newfloat{algorithm}{tbp}{loa}
\providecommand{\algorithmname}{Algorithm}
\floatname{algorithm}{\protect\algorithmname}

\usepackage{adjustbox}
\usepackage{multirow}
\usepackage{subfig}
\usepackage[margin=1in]{geometry}
\usepackage{algorithm,algpseudocode}
\usepackage[bottom]{footmisc}
\usepackage{amsmath}
\usepackage{babel}
\usepackage{authblk}

\let\oldmaketitle\maketitle
\renewcommand{\maketitle}{
\begin{singlespace}
  \oldmaketitle
\end{singlespace}}

\makeatother

\title{Estimating the Competitive Storage Model with Stochastic Trends in Commodity Prices}
\author[1]{Kjartan Kloster Osmundsen\thanks{Corresponding author. Email: kjartan.osmundsen@gmail.com}}
\author[1]{Tore Selland Kleppe}
\author[2]{Roman Liesenfeld}
\author[3]{Atle Oglend}
\affil[1]{Department of Mathematics and Physics, University of Stavanger, Norway}
\affil[2]{Institute of Econometrics and Statistics, University of Cologne, Germany}
\affil[3]{Department of Safety, Economics and Planning, University of Stavanger, Norway}

\begin{document}

\maketitle

\begin{abstract}
We propose a state-space model (SSM) for commodity prices that combines the competitive storage model with a stochastic trend.  This approach fits into the economic rationality of storage decisions, and adds to previous deterministic trend specifications of the storage model. Parameters are estimated using a particle Markov chain Monte Carlo procedure. Empirical application to four commodity markets shows that the stochastic trend SSM is favored over deterministic trend specifications.  The stochastic trend SSM identifies structural parameters that differ from those for deterministic trend specifications. In particular, the estimated price elasticities of demand are significantly larger under the stochastic trend SSM.\\
\noindent
\end{abstract}

\section{Introduction}
Economic theories are often developed in a stationary context. However, the real world does not always correspond to stationarity. This potential mismatch  creates  a challenge when attempting  to relate theory to historical data.
This  is a well-known problem in empirical macroeconomics, where structural parameters of business cycle models are often estimated on data that have been filtered in order to remove variation at frequencies  that the model is not intended to explain, such as low-frequency trend variations and seasonal fluctuations  \citep{deJong2007,sala2015dsge}. For an overview of alternatives to the use of pre-filtered data in order to address this general problem, see \citet{canova2014bridging}.

In the competitive storage model for commodity prices introduced by \citet{gustafson1958carryover}, the situation is similar to that of business cycle models.
The rational expectations equilibrium implied by the solution of this model is only known to exist in a stationary market.
Accordingly, it is a model for describing  dynamic price adjustments towards an exogenously given fixed
steady-state equilibrium.
However, it cannot explain low-frequency price movements due to persistent shocks.
This is problematic when attempting to estimate the structural parameters of the model using commodity price data,
since  time series of commodity prices typically display a strongly persistent behavior in the price level, so that non-stationarity cannot be rejected when using conventional statistical tests
(\citealp{wang2007commodity,Gouel2017}). As a result, the estimates for the structural parameters, which determine quantities like the price elasticity of demand and storage costs, are likely to be biased. This issue was recognized
by \citet{Deaton1995} in one of the earliest attempts to directly
estimate the structural parameters of the storage model.

This paper proposes an approach to estimate the structural parameters of the competitive commodity storage model using a state-space model (SSM) for commodity prices, which decomposes the observed price into a stationary component which is due to the storage model and a stochastic trend component included to capture low-frequency price variations the storage model is unable to explain.
Using a stochastic trend specification to account for non-stationary price data, our empirical approach aims at fitting into the economic rationality of the stationary storage model so that it  preserves theoretical coherence, promising meaningful estimates of the structural parameters.
Such a fit results from the fact that a stochastic trend that scales equilibrium prices can be isolated in the storage model by assuming that the innovations to the trend do not interfere with the agents' equilibrium  storage decisions.
In the baseline storage model, unrestricted equilibrium storage decisions  lead to an intertemporal pricing restriction of the form $P_{t}=\beta E_{t}\left(P_{t+1}\right)$, where $E_{t}(P_{t+1})$ is the rational period-$t$ expectation of the commodity price $P_{t+1}$ and $\beta$ represents  some discount factor. Thus, a stochastic price scaling $K_t$ will not impair the equilibrium storage decisions if  $K_{t}P_{t}=\beta E_{t}\left(K_{t+1}P_{t+1}\right)$. This generically identifies stochastic trends as  shifts in the price levels that do not interfere with intertemporal stock allocations, allowing a coherent  integration of  the stationary rational expectations equilibrium   into  a non-stationary environment, thus providing the theoretical basis of our empirical SSM approach.
The corresponding SSM, that jointly identifies the trend parameters and the structural parameters of the storage model, is non-linear in the latent states so that its likelihood function is not available in closed form.
To overcome this difficulty, we propose to use a Bayesian posterior analysis based on a particle Markov chain Monte Carlo (PMCMC) procedure \citep{andrieu_particle_2010}.

With our proposed approach we contribute to the literature concerned with the general problem of adapting stationary economic models to non-stationary data, and more specifically to the
problem of estimating the structural parameters of the competitive storage model on non-stationary commodity price data.  \citet{legrandempirical}
identifies reliable estimation as one of the main issues of structural models for commodity prices. Early attempts of estimating the structural parameters  revealed that  fitted competitive storage models are  not able to satisfactorily approximate the observed strong serial dependence in commodity price data, indicating misspecification of the empirical model and casting doubt on the reliability of the parameter estimates \citep{Deaton1995}.
Suggested solutions  to this problem include  ad-hoc  enrichments of the dynamic structure of the storage model by including weakly dependent supply shocks \citep{Deaton1996, kleppe_estimating_2017}, or the tuning of the grid for the commodity stock state variable, used for approximating the policy function \citep{cafiero2011empirical}. Other approaches  replace the estimation techniques applied in early empirical implementations of the storage model, like the pseudo maximum likelihood (ML) procedure of \cite{Deaton1996},  by more sophisticated ones, such as the ML technique developed by \citet{cafiero2015maximum} or the  particle filtering
methods proposed in \citet{kleppe_estimating_2017}.

Empirical approaches that, like ours, decompose the observed price into a component to be explained by the storage model and a trend component are those of  \citet{cafiero2011storage}, \citet{bobenrieth2013stocks}, \citet{guerra2015empirical} and \citet{Gouel2017}. The first three of these  studies  propose to account for the strong persistence in the price data that the storage model is not able to approximate, by detrending the prices using a deterministic log-linear trend prior to the estimation of the structural parameters. \citet{Gouel2017} improves upon this procedure by  jointly estimating the structural and deterministic trend parameters using the ML-estimator of \citet{cafiero2015maximum}.  The trend specifications \citet{Gouel2017} consider in their empirical application  include log-linear trends as well as more flexible trends specified as restricted cubic splines. One of their main findings is that empirical models accounting for a properly specified trend component in the observed commodity price yield more plausible estimates of the structural parameters than models without a trend.
However, the deterministic trends used in those studies inherently imply well predictable capital gains in the storage model, and so question the economic logic of separating the trend from structural economic pricing components.
Moreover, the appropriate functional form of the deterministic trend needs to be tailored  to the specific commodity market and the sampling frequency  for which the storage models are applied. In contrast, the stochastic trend as used in our SSM approach represents, in Bayesian terms,  a  hierarchical prior for the low-frequency price component, which is not only consistent with the rationality of the economic model, but also flexible in its design to account for variation that the storage model is not intended to explain. This makes our approach applicable to a broad range of commodity markets and different sampling frequencies. The strategy of scaling prices to address non-stationarity was also done
by \citet{Routledge2000} in their equilibrium term structure model
of crude oil futures. However, they did not do so in a rigorous estimation framework.

A stochastic trend as used in our storage SSM allows a potentially large fraction of the observed variation
in commodity prices to be accounted for by the trend component. This
risks miss-assigning price variation due to speculative storage to
the trend component.  Thus, if considered as an evaluation of the empirical
relevance of the storage model, the use of a stochastic trend can be considered as a conservative
test. To explore this issue further we perform a simulation experiment.
The simulation results suggest that our proposed approach is able to
accurately assign price variation to trend and model components.
We further apply our storage  SSM to  monthly observations of nominal coffee,
cotton, aluminum and natural gas prices. The results show, not surprisingly,
that most of the observed price variation is due to the stochastic
trend component.  In order to assess the empirical relevance of the competitive storage model, we compare the storage SSM to the nested model that results in the absence of storage.
The comparison reveals that the storage model predicting non-linear price dynamics with episodes of isolated  price spikes and increased volatility  adds significantly
to explaining the observed commodity price behavior.
We also compare the stochastic
trend SSM to the deterministic trend models of \citet{Gouel2017} by using the Bayes factor and a model residual analysis. Results
show that the SSM with stochastic trend fits the price data much better  than models with deterministic trends. The estimates for the price elasticity
of demand obtained from the stochastic trend SSM are substantially larger than  for the deterministic trend models. Also, the estimated storage costs vary considerably depending on the commodity.
This highlights the importance of properly accounting
for the trend behavior when evaluating the role of speculative storage
in commodity markets.

The rest of this paper is structured as follows. In the next section,
we present the storage model used in the paper and the assumed price
representation. We then present the estimation methodology (Section 3), simulation
results (Section 4) and empirical results for historical data (Section 5). We discuss the findings
before we offer some concluding remarks (Section 6).

\section{Storage Model}
\subsection{State-Space Formulation with Stochastic Trend}
Our approach relies upon the commodity storage model of \citet{oglend_behavior_2017}. It extends the \citet{deaton_behaviour_1992} model by including
an upper limit of storage capacity, $C\geq0$, in addition to the conventional non-negativity constraint for stocks, so that the  storage space is completely bounded. This upper limit takes into account possible congestion of the storage infrastructure, which can lead to negative price spikes in the event of substantial oversupply in the market. In addition, the assumption of a completely bounded storage space allows numerical solutions of the model that are more robust over a wider parameter range than those for a model without this assumption \citep{oglend_behavior_2017}. This, in turn,  simplifies estimation of the model parameters.

The economic model of commodity storage is a canonical  dynamic stochastic partial  equilibrium  model in discrete time for a commodity market with risk neural storage
agents and rational expectations. The rational expectations equilibrium is characterized by  a price function, denoted by $f(x)$, which maps the stocks  $x$ to commodity prices.
For empirical implementation, we assume that  the observed commodity price can be decomposed into a component to be explained by the commodity storage model  and a stochastic trend component. The corresponding time series model that we propose for the commodity log-price $p_{t}$, observed at time $t$ ($t=1,\ldots, T$), has the form
\begin{align}
p_{t} & =k_{t}+\log f(x_{t}), \label{eq:Storagemodel}\\
k_{t} & =k_{t-1}+\varepsilon_{t},\qquad\varepsilon_{t}\sim\text{iid }N(0,v^{2}), \label{eq:Storagemodel-2}\\
x_{t} & =(1-\delta)\sigma(x_{t-1})+z_{t},\qquad z_{t}\sim\text{iid }N(0,1)\label{eq:Storagemodel-3},
\end{align}
where the available quantity of commodity stocks $x_t$ is treated as  a latent state variable.
Its dynamics are linear in the equilibrium storage policy $\sigma(\cdot)$,
with stock depreciation rate $\delta$ and  Gaussian supply shocks $z_t$. The
latent trend component of the log-price $k_t$ is specified as
a driftless Gaussian random walk, so that it is allowed to vary gradually over time. The innovations of this stochastic trend $\varepsilon_{t}$ and the  supply shocks $z_{t}$ are assumed to be serially and  mutually independent.

The rational expectations equilibrium price function $f(x)$ satisfies for  all $x$
\begin{align}
f(x) & =\min\left\lbrace P(x-C),\max\left[ \bar{f}(x),P(x)\right] \right\rbrace ,\label{eq:fx}\\
\bar{f}(x) & =\beta\int f\Big(\left(1-\delta\right)\sigma(x)+z\Big)\phi(z)dz,\\
\sigma(x) & =x-D(f(x)),\label{eq:sig_f_rel}
\end{align}
where $D(p)$ represents a continuous and monotonically decreasing aggregate
demand function in the market, $P(x)$ is the corresponding inverse demand, and $\phi(z)$ is the  probability density function of the supply shock $z$.
The storage cost discount factor is given by $\beta=(1-\delta)/(1+r)$, where $r$ is a relevant
interest rate. According to Equation (\ref{eq:fx}), the equilibrium pricing function exhibits three different pricing regimes: (i)
a stock-out pricing regime, where $f(x)=P(x) \Leftrightarrow \sigma(x)=0$, (ii)
a no-arbitrage pricing regime, i.e.~$f(x)=\bar{f}(x) \Leftrightarrow C>\sigma(x)>0$,
where $\bar{f}(x)$ is the expected next period commodity price, and
(iii) a full capacity pricing regime, where  $f(x)=P(x-C)\Leftrightarrow\sigma(x)=C$.
The stock-out regime is characterized by positive price spiking and
high price volatility due to reduced shock buffering capabilities in
the market. Under the no-arbitrage regime, prices evolve smoothly with
a relatively low volatility. Full capacity pricing mirrors the stock-out
regime but with negative price spikes. As the market transitions between
regimes, prices move between periods of quiet and turmoil, generating non-linear dynamics in the price process. The rational
expectations equilibrium  $f(x)$ is stationary, having an associated globally
stationary price density \citep{oglend_behavior_2017}.

Using $k_{t}=p_{t-1}-\log f(x_{t-1})+\varepsilon_{t}$ in the price equation, the model as given in Equations (\ref{eq:Storagemodel})-(\ref{eq:Storagemodel-3}) can be written as
\begin{align}
p_{t} & =p_{t-1}+\log\left(\frac{f(x_{t})}{f(x_{t-1})}\right)+\varepsilon_{t},\qquad\varepsilon_{t}\sim\text{iid }N(0,v^{2}),\label{eq:p_p}\\
x_{t} & =(1-\delta)\sigma(x_{t-1})+z_{t},\qquad z_{t}\sim\text{iid }N(0,1).\label{eq:p_x}
\end{align}
This defines a non-linear Gaussian state-space model, with
measurement equation~(\ref{eq:p_p}) for the observed  price and state-transition equation~(\ref{eq:p_x}) for the latent stocks.

\subsection{Stochastic Trends and Storage Decisions}
Separating the trend from the storage model pricing component in a
consistent way that does not compromise the rationality of storage
agents in the market requires that the trend does not interfere with
intertemporal allocation incentives.
The martingale property of a trend component specified as a stochastic trend with innovations that are independent of supply shocks ensures that this requirement is met.
By using a separable stochastic trend
we are assuming that storage agents do not alter their storage decisions
based on trend innovations. In other words, trend innovations are
assumed perceived by agents as permanent scalings of price levels
that do not warrant adjustments to storage allocations.

As an example, consider permanent shocks $K$ to the inverse aggregate
demand in the market, $P^*=KP$. The aggregate demand implied by $P^*$ is $D^*$, and
the resulting
rational expectations equilibrium is given by
\begin{align}\label{eq:equilibrium_eq-1}
f^*(x)=\min\left\{ P^*(x-C),\max\left[ \beta\int f^*\Big(\left(1-\delta\right)\left(x-D^*\left(f^*(x)\right)\right)+z\Big)\phi(z)dz,P^*(x)\right] \right\} .
\end{align}
Assume the scaling process is given by $K'=\gamma K+\epsilon$, where
$\epsilon$ is a random variable with density $\phi_{\epsilon}$ which is independent of the supply shock $z$.
This scaling does not affect the optimal storage policy if $f^*(x)=Kf(x)$
solves the functional equation problem in Equation (\ref{eq:equilibrium_eq-1}), where $f(x)$ is the
rational expectations equilibrium for the original non-scaled prices.
Substituting the proposed solution for $f^*(x)$, we get
\begin{align}
&Kf(x) =  \min \Bigg\{ KP(x-C),\Bigg.\\
&\qquad\qquad\qquad\quad \left. \max\left[ \beta\int\left(\gamma K+\epsilon\right)\int f\Big((1-\delta)(x-D^*(Kf(x)))+z\Big)\phi(z)\phi_{\epsilon}(\epsilon)dzd\epsilon,KP(x)\right] \right\}.\nonumber
\end{align}
Note that $D^*(Kf(x))=D(f(x))$ by the definition of $D^*$
as the inverse of the scaled inverse demand function $P^*=KP$,
where $P=f(x)$. And so,
\begin{align}
&Kf(x)=\min \Bigg\{ KP(x-C),\Bigg.\\
&\qquad\qquad\qquad\quad \left. \max\left[ \beta\int\left(\gamma K+\epsilon\right)\int f\Big((1-\delta)(x-D(f(x)))+z\Big)\phi(z)\phi_{\epsilon}(\epsilon)dzd\epsilon,KP(x)\right] \right\}\nonumber.
\end{align}
If $\int\left(\gamma K+\epsilon\right)\phi_{\epsilon}(\epsilon)d\epsilon=K$ implying that $E(K')=K$,
we obtain
\begin{align}
Kf(x)=K\min\left\{ P(x-C),\max\left[ \beta\int f\Big((1-\delta)(x-D(f(x)))+z\Big)\phi(z)dz,P(x)\right] \right\},
\end{align}
which establishes $f^*(x)=Kf(x)$ as the solution to the inverse
demand scaled rational expectations equilibrium. Consequently, any
observed commodity price can be represented as $P=Kf(x)$. Formally,
the rational expectations equilibrium is linear homogeneous to the
proportional scaling $K$ of the inverse aggregate demand function
when $E(K')=K$ and innovations to the trend are orthogonal to supply
shocks.

Note that in our econometric model as given by Equations (\ref{eq:Storagemodel})-(\ref{eq:Storagemodel-3}), it is the logarithm of the scaling process for the price levels $k=\log(K)$ and not $K$, for which we assume a
stochastic trend. Hence, the martingale property for the scaling term process $K$ will not apply exactly, and the assumed Gaussian  process for $\log(K)$ implies that $E(K')=K\exp\left(v^{2}/2\right)>K$.
By ignoring this bias in our econometric model we make the behavioral assumption
that agents do not alter storage decision based on the capital gain due to the expected mark-up factor
$\exp\left(v^{2}/2\right)>1$.
We consider this a reasonable trade-off to allow us to empirically analyze the storage model within  a log-linear
state and measurement space, which is comparatively convenient  for statistical inference.
In addition, the bias is small when $v^{2}$ is small, that is, when the trend is fairly smooth. In fact, the estimates we obtain for $v$  in our empirical application discussed below imply that the factor $\exp\left(v^{2}/2\right)$ varies in a range between 1.001 and 1.004 so that it is essentially negligible. Ignoring this factor is
essentially equivalent to transforming the probability
space to a setting where agents ignore information from trend innovations,
similar to a risk-neutral valuation setting where $v^{2}$ defines
a required risk premium term or a nominal inflation term.\\

\section{Statistical Inference}
\subsection{Preliminaries}
In our empirical application of the storage model with stochastic trend based on its state-space representation as given in Equations (\ref{eq:p_p}) and (\ref{eq:p_x}), we use monthly commodity spot prices and rely on a Bayesian Markov chain Monte Carlo (MCMC) posterior analysis. For this application, we follow \citet{kleppe_estimating_2017} and use $P(x)=\exp(-bx)$ as the inverse demand function, where the parameter $b$ measures the  semi-elasticity of the demand price. In line with \cite{Gouel2017}, we fix the yearly interest rate at 5\%, so that the monthly storage cost discount factor is given by $\beta=(1-\delta)/(1+r)$, with $r=1.05^{1/12}-1$. The set of parameters then consists of the structural parameters $(\delta,b,C)$ and the trend parameter $v$.

In initial experiments to estimate the parameters, we found that the capacity limit $C$ is empirically not well identified separately from the remaining parameters. This appears to be mainly due to the fairly small sample size of our data, ranging from 264 to 360 monthly spot price observations. Therefore, we decided to fix $C$ at a positive predetermined value. Since $C$  determines the full capacity threshold for equilibrium storage, it bounds the space for the unit-free latent state variable $x$, and  by fixing its value (together with  normalizing the mean of $z$ to zero) we pin down the range of this space. The values of the remaining parameters $(\delta, b,v)$ and their implications for the price dynamics are then to be interpreted relative to this scale of $x$. In our empirical application below we set the capacity limit $C=10$. This ensures that it is a fairly  rare event for the market to be in the full-capacity regime. Suppose, for example,  that all realizations of the supply shocks $z$ for a sequence of periods are equal to one standard deviation and that nothing is consumed in those periods, so that $x_{t+1}=(1-\delta)x_t+1$. Then  a storage infrastructure  with $C=10$ and a notable depreciation rate of $\delta=0.01$ can store those unconsumed supplies for about 10 months before  reaching the capacity  limit\footnote{Our selection of $C=10$ also corresponds to the lowest upper limit, which  \citet[Table I]{Deaton1995} use for their grids of $x$-values in the interpolation scheme to compute the equilibrium price function for a set of various yearly commodity prices.  The upper  grid boundaries for the different commodities have been chosen by the authors so that the calculations never generate $x$-values that exceed these maximum values for the grid.}.

In order to solve the functional equation for the equilibrium price function $f(x)$ as defined by Equations (\ref{eq:fx})-(\ref{eq:sig_f_rel}), we use a numerical algorithm which is based on the method of \citet{oglend2019}, detailed in Appendix \ref{subsec:Price-function}.
This algorithm takes advantage of the fact that the storage space is completely bounded by the non-negative constraint and the capacity limit $C$, thus providing numerically robust and computationally fast solutions. This is critical for a Bayesian MCMC posterior analysis because it requires a significant number of reruns of the algorithm to obtain a solution to the pricing function for each new parameter value.

\subsection{Bayesian Inference Using Particle Markov Chain Monte Carlo}
In the SSM model as given by  Equations (\ref{eq:p_p}) and (\ref{eq:p_x}) the vector of parameters to be estimated is given by  $\theta=(v,\delta,b)$ and the vector of latent state variables is $x_{1:T}$, where  the notation $a_{s:s'}$ is used to denote $(a_s,a_{s+1},\ldots,a_{s'})$. The posterior of the parameters is $\pi(\theta|p_{1:T})\propto \pi_\theta(p_{1:T}) \pi(\theta)$,
where $\pi(\theta)$ denotes the prior density assigned to $\theta$ and $\pi_\theta(p_{1:T})$ represents the likelihood function, given by
\begin{equation}
\pi_\theta(p_{1:T})=\int\left[\prod_{t=2}^T \pi_\theta(p_t|p_{t-1},x_{t-1:t})\pi_\theta(x_t|x_{t-1}) \right]\pi_\theta(p_1,x_1) dx_{1:T} ,\label{eq:likelihood}
\end{equation}
with
\begin{align}
 \pi_\theta(p_t|p_{t-1},x_{t-1:t})={\cal N}\left(p_t|p_{t-1}+\log\left(\frac{f(x_t)}{f(x_{t-1})}\right),v^2\right),\quad
 \pi_\theta(x_t|x_{t-1})= {\cal N}\left(x_t| (1-\delta)\sigma(x_{t-1}),1\right),\label{eq:p_px}
\end{align}
where ${\cal N}(\cdot|\mu,\sigma^2)$ denotes a normal density function with mean $\mu$ and variance $\sigma^2$. For the joint density  of the price and state in the initial period $\pi_\theta(p_1,x_1)$ we assume that it factorizes into a uniform density on $(-2,C+2)$ for the state $x_1$, denoted by ${\cal U}(x_1|-2,C+2)$, and a dirac measure for the price $p_1$ located at its actually observed value (effectively conditioning the likelihood on the first price observation).

Due to the non-linear nature of the  pricing function $f(x)$ and the storage function $\sigma(x)$  entering the measurement and state transition density as given in Equation (\ref{eq:p_px}), the likelihood (and hence the resulting posterior for $\theta$) are not available in closed form, so that a Bayesian and  likelihood-based inference  requires approximation techniques.
Several Monte Carlo (MC) approximation approaches have been developed for statistical inference in non-linear SSMs with analytically intractable likelihood functions.
However, only a few of them are suited to the model considered here due to the discontinuous  derivatives of $f(x)$. In particular,  methods using MC estimators for the likelihood  $\pi_\theta(p_{1:T})$ based on approximations to the conditional posterior  of the states $\pi(x_{1:T}|\theta, p_{1:T})$, including  second order/Laplace approximations \citep{shepard_pitt97,durbin_koopman_2ed} or  global approximations  as used by  the efficient importance sampler \citep{liesenfeld_richard_03,richardetal07}, perform poorly in such a context.
The same applies to the Gibbs approach targeting the joint posterior distribution of the states and parameters $\pi(x_{1:T},\theta|p_{1:T})$ and alternately simulating from the conditional posteriors $\pi(x_{1:T}|\theta,p_{1:T})$ and $\pi(\theta |x_{1:T},p_{1:T})$. It is known that such a Gibbs procedure typically has problems in efficiently approximating the targeted joint posterior in non-linear SSMs due to a fairly slow mixing \citep{BosShephard2006}.
Moreover, the Gibbs procedure is also not very computationally attractive in the present context, since both the (joint) conditional posterior of all the states  $\pi(x_{1:T}|\theta, p_{1:T})$ and the single-site conditional posterior of the individual states $\pi(x_{t}|x_{1:t-1}, x_{t+1:T} ,\theta, p_{1:T})$ are non-standard distributions.

Here, we propose to use the particle marginal Metropolis-Hastings (PMMH) approach as developed by \cite{andrieu_particle_2010}, which is well suited for a posterior analysis of our proposed storage  SSM as it can cope with the discontinuity of the gradients of $f(x)$ and is very easy to implement. The PMMH uses unbiased MC estimates of the likelihood $\pi_\theta(p_{1:T})$ inside a standard Metropolis-Hastings (MH) algorithm  targeting the posterior of the parameters $\pi(\theta|p_{1:T})$.
The MC estimation error of the likelihood estimate does not affect the invariant distribution of the MH so that the PMMH allows for exact inference. The PMMH produces an MCMC sample $\{ \theta_i\}_{i=1}^S$ from the target distribution by the following MH updating scheme: Given the previously sampled $\theta_{i-1}$ and the corresponding  likelihood estimate  $\hat \pi_{\theta_{i-1}}(p_{1:T})$, a candidate  value $\theta_{*}$ is drawn from a proposal density $Q(\theta|\theta_{i-1})$, and the estimate of the associated  likelihood $\hat \pi_{\theta_{*}}(p_{1:T})$ is computed.
Then the candidate $\theta_{*}$ is accepted as the next simulated  $\theta_i$ with probability
\begin{equation}
\alpha(\theta_{*},\theta_{i-1})=\min\left\{ 1,\frac{\hat{\pi}_{\theta_{*}}(p_{1:T})\pi(\theta_{*})}{\hat{\pi}_{\theta_{i-1}}(p_{1:T})\pi(\theta_{i-1})}\frac{Q(\theta_{i-1}|\theta_{*})}{Q(\theta_{*}|\theta_{i-1})}\right\},\label{eq:MHacc}
\end{equation}
otherwise $\theta_i$ is set equal to $\theta_{i-1}$. Under weak regularity conditions, the resulting sequence $\{ \theta_i\}_{i=1}^S$ converges to samples from the target density  $\pi(\theta|p_{1:T})$ as $S\to\infty$
\citep[][Theorem 4]{andrieu_particle_2010}.

For the  PMMH, we use a Gaussian random walk proposal density $Q(\theta|\theta_{i-1})={\cal N}(\theta|\theta_{i-1},\Sigma)$ and follow  the
approach of \cite{haario_adaptive_2001} to adaptively set the proposal covariance matrix $\Sigma$ during the burn-in period of the MCMC iterations.
After dropping the draws from the burn-in period, we use the $\theta$ draws from the next $M$ PMMH iterations to represent the posterior $\pi(\theta|x_{1:T})$. The posterior mean of the parameters, used as point estimates, is approximated by the sample mean over the $M$ PMMH draws.
For numerical stability of the PMMH computations, we reparameterize the likelihood function using the transformed parameters $\theta=\big(\log(v),\text{arctanh}(2\delta-1), \log(b)\big)$ so that the resulting parameter space is unconstrained.

The prior densities for the parameters are selected as follows:  For $\log(b)$ we assume a $N(0,1)$ prior, and for $v^2$ an inverted chi-squared prior with  $v^2\sim 0.1/\chi_{(10)}^2$, where $\chi_{(10)}^2$ denotes a chi-squared distribution with 10 degrees of freedom. Under this prior for $v^2$, the mean is  given by 0.01 and the standard deviation by 0.007.
The prior density assigned to $\delta$ is a Beta with $\delta\sim {\cal B}(2,20)$ so that the  mean and standard deviation are given by  0.09 and 0.05, respectively.

\subsection{Particle Filter Likelihood Evaluation}\label{sec:BPF}
In  order to obtain unbiased MC estimates for the likelihood in Equation (\ref{eq:likelihood}), required as an input of the PMMH, we follow \cite{andrieu_particle_2010} and \cite{flury_shephart_2011} and use  a simple sampling importance resampling
(SIR) particle filter (PF).  For  given values of the parameters $\theta$, it produces  MC estimates for the sequence of period-$t$ likelihood contributions $\pi_\theta(p_t|p_{1:t-1})$ by sequentially sampling and resampling using an   importance  sampling (IS) density $q(x_t|x_{1:t-1})$ for the states $x_t$ \citep[see,][for a detailed treatment of PFs]{doucet_tutorial_2009,Cappe2007}. For the implementation of the PF we use the state-transition density $\pi_\theta(x_t|x_{t-1})$ as IS density   \citep{Gordon1993}, and rely on a dynamic resampling scheme in which the particles are resampled only when their effective sample size  falls below one half of the number of particles \citep{doucet_tutorial_2009}. This simple version of the PF (also known as the bootstrap PF, BPF) for approximating the likelihood as given by  Equations (\ref{eq:likelihood}) and (\ref{eq:p_px}) consists of the following steps:

{\it For period $t=1$ (initialization)}: Sample $x_1^k\sim\pi_1(x_1)={\cal U}(x_1|-2,C+2)$ for $k=1,\ldots,N$ and set the corresponding (normalized)  IS weights to $W_1^k=1/N$.  
For initialization set $\bar x_1^k= x_1^k$.

{\it For periods $t=2,\ldots, T$}: Sample $x_t^k\sim\pi_\theta(x_t| \bar x_{t-1}^k)={\cal N}\big(x_t| (1-\delta)\sigma(\bar x_{t-1}^k),1\big)$ for $k=1,\ldots,N$ and set $x^k_{1:t}=(x_t^k,\bar x_{1:t-1}^k)$. Compute the IS weights as
\begin{align}
w_t^k = W^k_{t-1}\pi_\theta(p_t|p_{t-1},x_{t-1:t}^k),
\end{align}
and their normalized versions  $W_t^k=w_t^k/(\sum_{\ell=1}^N w_t^\ell)$. Then use  the IS weights   to obtain the period-$t$ likelihood contribution as $\hat\pi_\theta(p_t|p_{1:t-1})= (\sum_{k=1}^N w_t^k)/N$, and compute the effective particle sample size defined by $N^e_t=[\sum_{k=1}^N (W_t^k)^2]^{-1}$. If $N^e_t<N/2$, resample from the  particles $\{x_{1:t}^k\}_{k=1}^N$ with replacement according to their IS weights $W_t^k$ to obtain the resampled particles $\{\bar x_{1:t}^k\}_{k=1}^N$, and set their weights to $W_t^k=1/N$. Otherwise, set $\bar x_{1:t}^k=x_{1:t}^k$.

The resulting BPF estimate for the likelihood (conditional on the first price observation) is given by $\hat \pi_\theta(p_{1:T})=[\prod_{t=2}^T \hat\pi_\theta(p_t|p_{1:t-1})]$.
The measurement density $\pi_\theta(p_t|p_{t-1},x_{t-1:t})$ is not very informative about the  states $x_t$ for empirically relevant parameter values, resulting in a low signal-to-noise ratio. Thus, the simple BPF yields fairly precise  MC estimates of the likelihood with a modest number of particles $N$ \citep{Cappe2007}. High precision likelihood estimates are a critical requirement for the PMMH to produce a well mixing MCMC sample from the posterior of the parameters $\pi(\theta|x_{1:T})$ \citep{flury_shephart_2011}.
In our applications below, we use $N=10,000$ particles. For a time series with
$T=360$, one BPF likelihood estimate requires approximately 2.5 seconds (on a computer
with an Intel Core i5-6500 processor running at 3.20 GHz). The MC numerical standard deviation of the log-likelihood estimate  $\log\hat{\pi}_\theta(p_{1:T})$, computed from reruns of the BPF for a fixed $\theta$ value under different seeds, is about 0.1 percent of the absolute value of the log-likelihood, illustrating the high accuracy of the BPF.

\subsection{State Prediction and Diagnostics}
The BPF outlined in the previous section, and used for the PMMH implementation, can also be used to produce MC estimates for the predicted values of the latent state vector $x_{1:t+1}$ and functions thereof, given the prices observed up to period $t$, $p_{1:t}$. MC estimates of such predictions can serve as the basis for diagnostic checks. Let $h(x_{1:t+1})$ be a function of interest in $x_{1:t+1}$. Its conditional mean given $p_{1:t}$ can be expressed as
\begin{align}\label{eq:h_mean}
E\big(h(x_{1:t+1})|p_{1:t}\big)=\int h(x_{1:t+1})\pi_\theta(x_{t+1}|x_t) \pi_\theta(x_{1:t}|p_{1:t})dx_{1:t+1},
\end{align}
where $\pi_\theta(x_{t+1}|x_t)$ is the state-transition density as given by Equation (\ref{eq:p_px})  and $ \pi_\theta(x_{1:t}|p_{1:t})$ is the filtering density for $x_{1:t}$. Since the particles and IS-weights   $\{x_{1:t}^k,W_t^k\}_{k=1}^N$ produced by the BPF provide an MC approximation to this filtering density, the conditional mean in Equation (\ref{eq:h_mean}) for a given value of $\theta$ can be easily estimated by
\begin{align}\label{eq:h_mean_est}
\hat E(h(x_{1:t+1})|p_{1:t})=\sum_{k=1}^N h(x_{1:t+1}^k)W_t^k,
\end{align}
with $x_{1:t+1}^k=(x_{1:t}^k,x_{t+1}^k)$, where $x_{t+1}^k$ is obtained by propagating the BPF particle $x_{1:t}^k$ via the state-transition density, i.e. $x_{t+1}^k\sim \pi_\theta(x_{t+1}|x_t^k)$. In practice,  the parameters $\theta$ are set equal to their estimates.

This MC approximation of a predicted mean like that in Equation (\ref{eq:h_mean})  enables us to compute several useful statistics, such as the filtered mean  for the price function of the storage model $E(\log f(x_t)|p_{1:t})$  and the stochastic trend component $E(k_t|p_{1:t})=p_t-E(\log f(x_t)|p_{1:t})$, for which the function $h$ to be used is   $h(x_{1:t+1})=\log f(x_t)$.
State predictions can also be used to compute standardized Pearson residuals defined as
\begin{align}\label{eq:Pearson}
\eta_{t+1}=[p_{t+1}-E(p_{t+1}|p_{1:t}) ]/\text{Var}(p_{t+1}|p_{1:t})^{1/2}.
\end{align}
If the model is correctly specified, then $\eta_{t+1}$ and $\eta_{t+1}^2$ are serially uncorrelated so that they can be used for diagnostic checking of the assumed dynamic structure. The conditional moments of $p_{t+1}$ for the storage SSM are given by $E(p_{t+1}|p_{1:t})=p_t+E(\log[f(x_{t+1})/f(x_{t})]|p_{1:t})$ and $\text{Var}(p_{t+1}|p_{1:t})=\text{Var}(\log[f(x_{t+1})/f(x_{t})]|p_{1:t})+v^2$, which can be evaluated by Equation (\ref{eq:h_mean_est}), using the functions
$h(x_{1:t+1})=\log[f(x_{t+1})/f(x_{t})]$ and $h(x_{1:t+1})=\{\log[f(x_{t+1})/f(x_{t})]- \hat E(\log[f(x_{t+1})/f(x_{t})]|p_{1:t})\}^2$.

In order to check the capability of the storage SSM to approximate the distributional properties of the observed prices we use the probability integral transformed (PIT) residuals defined as
\begin{align}\label{eq:PIT}
\xi_{t+1}=\Phi^{-1}(u_{t+1}),\qquad u_{t+1}=\mbox{Pr}(p_{t+1}\leq p_{t+1}^{o}|p_{1:t}),
\end{align}
where $\mbox{Pr}(p_{t+1}\leq p_{t+1}^{o}|p_{1:t})$ is the predicted probability  that $p_{t+1}$ is less or equal to the actually ex-post observed price $p_{t+1}^{o}$, and $\Phi$ denotes the cdf of a $N(0,1)$-distribution \citep{KimShephardChib1998}. The PIT residuals $\xi_{t+1}$ follow a $N(0,1)$-distribution if the model is valid. For the storage SSM, the probability $u_{t+1}$ can be  calculated by setting the function $h(x_{1:t+1})$ equal to
$\Phi( \{p_{t+1}^{o}-  p_t - \log[f(x_{t+1})/f(x_{t})]\}/v)$.

\subsection{Marginal Likelihood for Model Comparison}
Marginal likelihood is used to compare the storage SSM with alternative models and assess the empirical relevance of the structural storage model component in the SSM. In order to evaluate the marginal likelihood  for the storage SSM we rely upon the procedure proposed by  \cite{ChibJeliazkov2001}, which is specifically customized for Bayesian analyses implemented using MH algorithms targeting the posterior of the parameters. This procedure takes advantage of the fact that the marginal likelihood can be expressed as
\begin{align}\label{eq:marg_lik}
\pi(p_{1:T})=\frac{\pi_{\bar\theta}(p_{1:T})\pi(\bar\theta) }{\pi(\bar\theta|p_{1:T})},
\end{align}
where  $\pi_{\bar\theta}(p_{1:T})$ is the likelihood function for the observed prices evaluated at some value of the parameters $\bar\theta$, and $\pi(\bar\theta) $ and  $\pi(\bar\theta|p_{1:T})$ are the corresponding ordinates of the prior and posterior of the parameters. Then it exploits that the posterior ordinate  $\pi(\bar\theta|p_{1:T})$ can be expressed in terms of the MH acceptance probability $\alpha(\cdot,\cdot)$ and the proposal density $Q(\cdot|\cdot)$. Namely,
as the ratio of the expectation of $\alpha(\bar\theta,\theta)Q(\bar\theta|\theta)$ under the posterior $\pi(\theta|p_{1:T})$ relative to the expectation of  $\alpha(\theta,\bar \theta)$ under the proposal density $Q(\theta|\bar\theta)$. This implies that a consistent MC estimate for $\pi(\bar\theta|p_{1:T})$ based on the MH acceptance probability defined in Equation (\ref{eq:MHacc}) is given by
\begin{align}
\hat\pi(\bar\theta|p_{1:T})=\frac{M^{-1}\sum_{i=1}^{M} \alpha(\bar\theta,\theta_i)Q(\bar\theta|\theta_i)}{L^{-1}\sum_{l=1}^{L} \alpha(\theta_l,\bar \theta)},
\end{align}
where $\{\theta_i\}_{i=1}^{M}$ are the $M$ simulated draws from the posterior distribution $\pi(\theta|p_{1:T})$ and $\{\theta_l\}_{l=1}^{L}$ are draws from the proposal distribution $Q(\theta|\bar\theta)$. For evaluating the likelihood $\pi_{\bar\theta}(p_{1:T})$ in Equation (\ref{eq:marg_lik}) we use the same BPF algorithm  as applied for the computation of the  MH acceptance probabilities in Equation (\ref{eq:MHacc}) (and outlined in Section \ref{sec:BPF}). The value of the point $\bar \theta$ is set equal to the posterior mean of $\theta$.

\section{Ability to Isolate the Trend and Storage Model Component}
In order to illustrate the capability of our Bayesian storage SSM approach to empirically separate  the variation in the observed prices into the variation   generated by the structural storage model component and that of the stochastic trend,  we conduct a simulation experiment. Prices are simulated from the storage SSM for parameters that are set equal to their posterior mean values found for the empirical application to natural gas prices,  discussed further below (see Table \ref{tab:pars}). Prices are simulated for 800 periods, with the first 500 discarded as burn-in, so that the size of the simulated sample is $T=300$.
The storage SSM is then fitted to the time series of simulated prices by using the PMMH procedure, and the BPF  is applied to produce estimates of the filtered mean for the  storage model price component $E(f(x_t)|p_{1:t})$ and the stochastic trend $E(k_t|p_{1:t})$, evaluated at the posterior mean of the parameters.
\begin{figure}[h]
\begin{centering}
\includegraphics[scale=0.45]{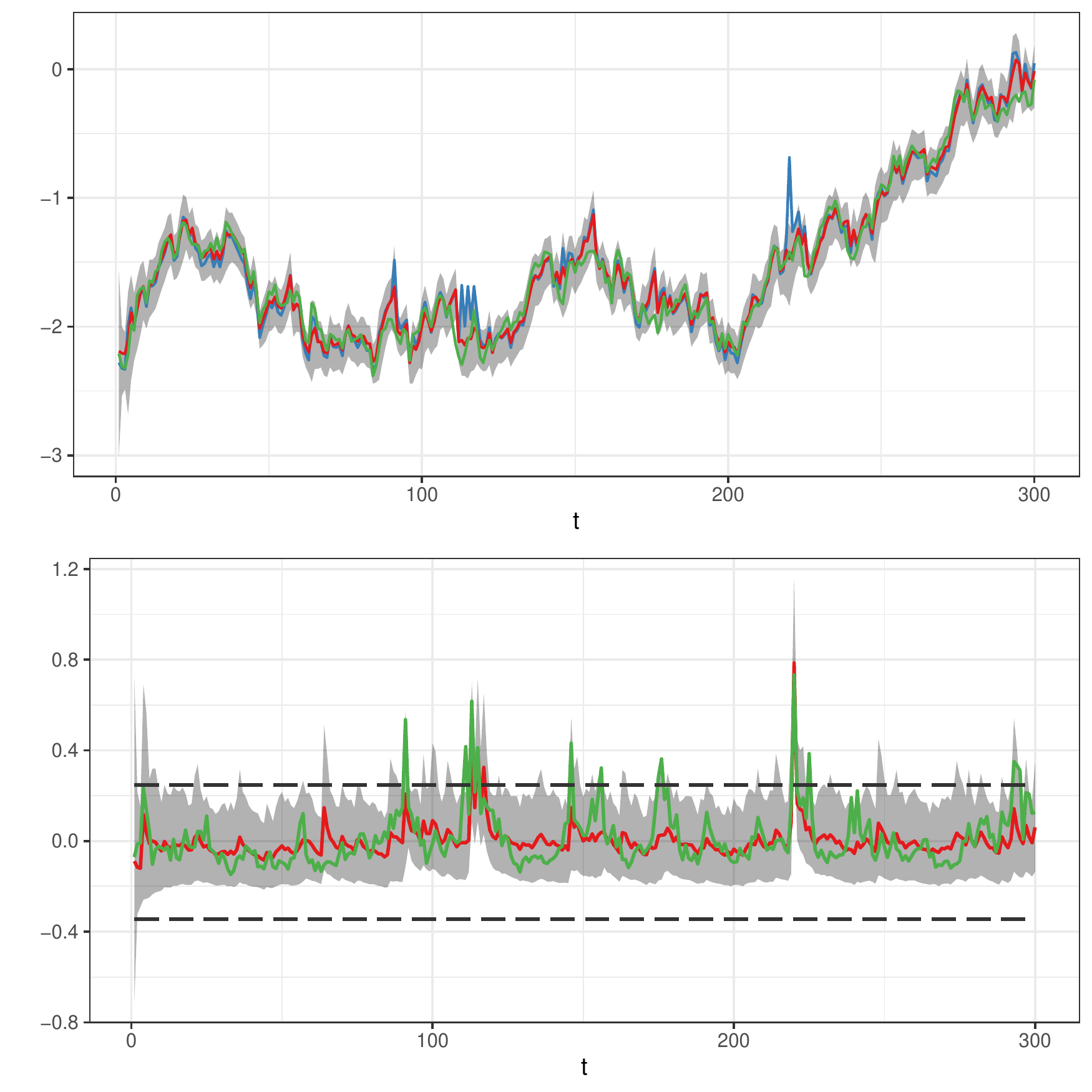}
\par\end{centering}
\caption{Filtered price components for simulated data. Upper panel: time series plot of the simulated log price $\log p_t$ (blue line), the actual stochastic trend component $k_t$ (green line), and its estimated filtered mean $E(k_t|p_{1:t})$ (red line). Lower panel: time series plot of the actual storage model component $\log f(x_t)$ (green line) and its estimated filtered mean $E(\log f(x_t)|p_{1:t})$ (red line). The gray shaded areas indicate the 95\%
credible intervals under the filtering densities for $k_t$ and $\log f(x_t)$, and the dashed lines in the lower panel mark the boundaries of the storage regimes. The prices are simulated using  parameters set at $(v,\delta,b)=(0.097,0.011, 0.420)$. The posterior mean of the parameters obtained by fitting the model to the simulated price data are $(\hat v,\hat \delta,\hat b)=(0.101,0.013, 0.436)$.  }
\label{fig:Simulation-experiment}
\end{figure}

Figure \ref{fig:Simulation-experiment} shows the results  of the simulation experiment. The upper panel displays the time series of the simulated log price $\log p_t$ and the actual simulated stochastic trend component $k_t$ together with the estimate of its filtered mean,
and  the lower panel shows the time series of the actual price component which is generated by the competitive storage model $\log f(x_t)$ together with its estimated filtered mean.
Also plotted are the boundaries of the storage regimes. Values of the price component
above the upper boundary correspond to the stock-out pricing regime, and values below the
lower boundary correspond to the full storage capacity  pricing regime.
The plotted time series show that the estimated  filtered means of the two price components track the time evolution of the true components quite well. We also observe that the estimated  filtered means  predict  most of the stock-out events, which is critically important for identifying the structural parameters of the storage model.
These simulation results illustrate that our approach appears to be well capable to empirically identify - from observed commodity prices - the fraction of the price variation which is due competitive storage decisions and to separate it from stochastic trend variation.

\section{Empirical Application}

In this section, we apply our Bayesian storage SSM approach to historical monthly price data for the following four commodities: Coffee ({\sl Coffee, Other Mild Arabicas, New York
cash price, ex-dock New York, US cents per pound}), cotton ({\sl Average
Spot Price in US cents per Pound for Upland cotton -- color 41, leaf
4, staple 34}), aluminum ({\sl Aluminum (LME) London Metal Exchange, unalloyed
primary ingots, high grade, minimum 99.7\% purity, USD per Metric
Ton}), and   natural gas ({\sl Natural Gas (U.S.), spot price at Henry Hub, Louisiana, USD per MBtu}). The respective sample periods range from Jan 1989 until Dec 2018 ($T=360$) for coffee, cotton and aluminum, and from Jan 1997 until Dec 2018 ($T=264$) for natural gas. All prices are in nominal
terms. We use monthly instead of annual prices to allow for more information about short-term price movements, as well as to avoid
potentially spurious averaging effects of annual prices \citep{guerra2015empirical}.

\subsection{Estimation Results for the Storage SSM with Stochastic Trend}

\begin{table}
\centering 
 \resizebox{0.55\textwidth}{!}{ 
\begin{tabular}{llllll}
   \hline
  &   & Natgas & Coffee & Cotton & Aluminum \\ 
   \hline
$v$ & Post. mean & 0.0972 & 0.0607 & 0.0459 & 0.0448 \\ 
    & Post. std. & 0.0083 & 0.0034 & 0.0027 & 0.0023 \\ 
    & ESS & 634 & 442 & 1028 & 990 \\ 
    &   &   &   &   &   \\ 
  $\delta$ & Post. mean & 0.0112 & 0.0023 & 0.0013 & 0.0011 \\ 
    & Post. std. & 0.0048 & 0.0013 & 0.0008 & 0.0007 \\ 
    & ESS & 580 & 376 & 761 & 847 \\ 
    &   &   &   &   &   \\ 
  $b$ & Post. mean & 0.4196 & 0.3849 & 0.3247 & 0.1987 \\ 
    & Post. std. & 0.2594 & 0.0913 & 0.0598 & 0.0676 \\ 
    & ESS & 515 & 386 & 972 & 917 \\ 
   \hline
\end{tabular}
 }

\caption{MCMC posterior analysis of the storage SSM with stochastic trend. The reported numbers are the posterior mean, posterior standard deviation and  effective sample size (ESS) for the parameters. The results are based on $12,000$ PMMH iterations, discarding the first 2000 burn-in iterations.}
\label{tab:pars}
\end{table}

For the Bayesian posterior analysis   of the storage SSM,  we run the PMMH algorithm for 12,000 iterations and discard the first 2,000 as burn-in.  In order to evaluate the sampling efficiency of the PMMH for estimating the parameters, we compute the effective sample size (ESS) of their posterior PMMH samples \citep{geyer1992}. The ESS measures the size of a  hypothetical  independent sample directly drawn from the posterior of the parameters which delivers the same numerical precision  as the actual sample of $M$  correlated PMMH parameter draws, so that large ESS values are to be preferred.

For each of the four commodities, the estimated posterior mean, standard deviation and ESS for the parameters are found in Table \ref{tab:pars}. The ESS values range from 376 to 1,028, indicating a satisfactory sampling efficiency with a fairly fast mixing rate of the PMMH algorithm. The estimates for the standard deviation of the trend innovations $v$ imply that the stochastic trend accounts for 53\% of the variation observed in the monthly price changes for natural gas, 66\% for coffee, 71\% for cotton, and 81\% for aluminum. As for the estimates of the depreciation rate $\delta$,  we observe that they are fully in line with the actual storage costs to be expected for the different types of commodities: For natural gas we find the largest estimated depreciation rate (1.1\%), which implies that the monthly cost of storage amounts to 1.5\% of the price.
This relatively large estimated storage cost is in accordance with the fairly expensive storage technology for US natural gas, which is typically stored in underground salt caves and  similar facilities. The second largest storage cost is found for coffee, with a monthly depreciation rate of 0.2\% leading to estimated monthly costs of 0.6\% of the price. The lowest storage costs are predicted for the non-food and non-energy products cotton and aluminum, for which the estimated depreciation rate is 0.1\% resulting in  storage costs of 0.5\%. We also observe that the larger the estimated storage cost for a commodity, the larger the fraction of observed price variation which is captured by the storage decision behavior. This is  in agreement with the rationality of the competitive  storage model, where higher storage costs are associated with more frequent stock-out events, which in turn  implies greater price volatility.
The posterior mean values for the slope parameter $b$ of the inverse demand function imply that a reduction in supply on the market by one standard deviation of production leads to a price increase of 42\% for natural gas, 38\% for coffee, 32\% for cotton and 20\% for aluminum.  The size of these estimated price elasticities roughly corresponds to the size of the price peaks observed in these markets.

\begin{figure}[h!]
\begin{centering}
\includegraphics[scale=0.4]{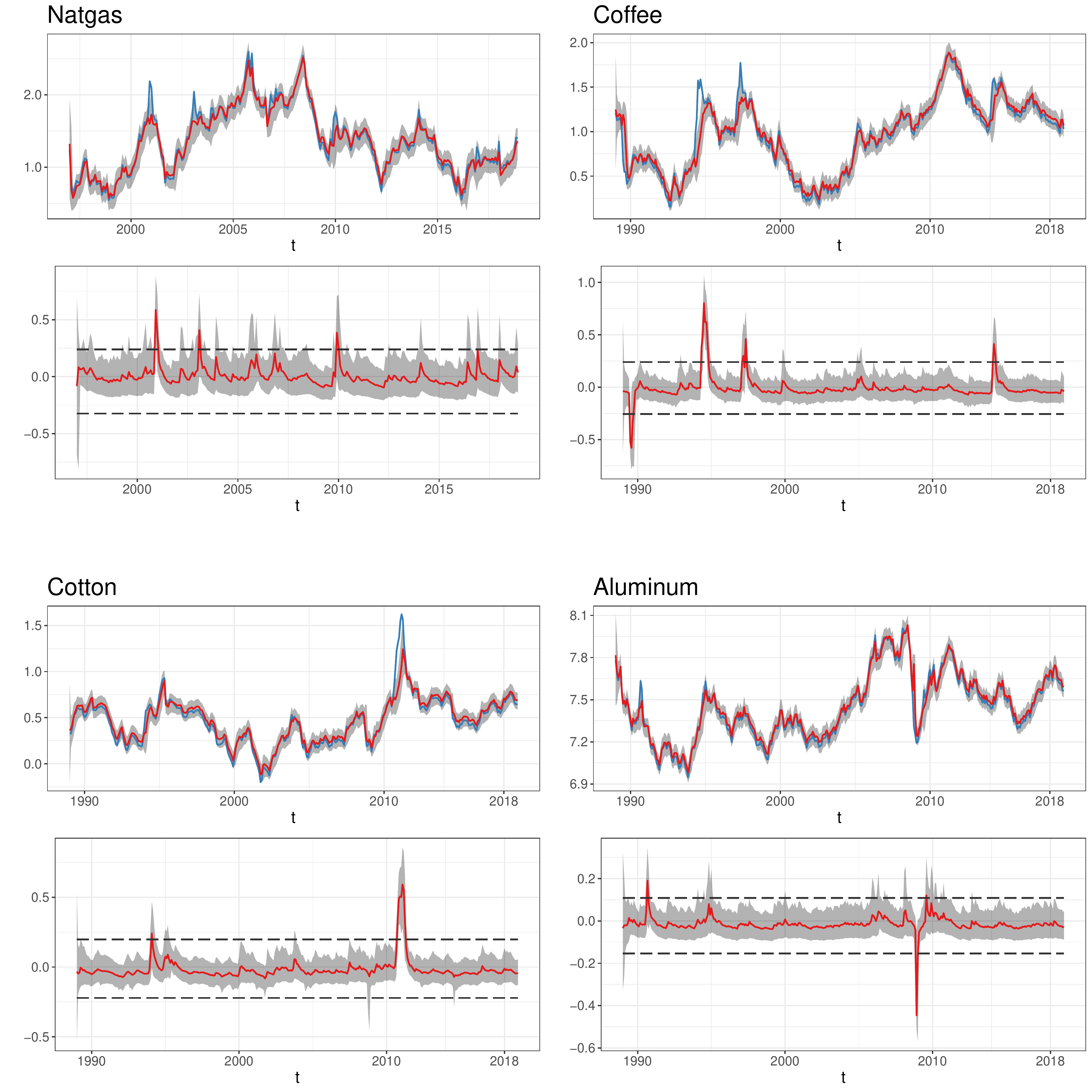}
\par\end{centering}
\caption{Commodity prices and filtered price components. Upper panels: time series plot of the log price $\log p_t$ (blue line) and the estimated filtered mean of the stochastic trend component $E(k_t|p_{1:t})$ (red line). Lower panels: time series plot of  the estimated filtered mean of the  storage model component $E(\log f(x_t)|p_{1:t})$ (red line). The gray shaded areas indicate the 95\% credible intervals under the filtering densities for $k_t$ and $\log f(x_t)$, and the dashed lines in the lower panels mark the boundaries of the storage regimes.
The parameters are set to their posterior mean as given in Table \ref{tab:pars}.\label{fig:filteredPriceComponents}}
\end{figure}

Figure \ref{fig:filteredPriceComponents} displays the time series of the log-prices for each of the four commodities, together with the filtered mean for their stochastic trend component  $k_t$ and their price component associated with the competitive storage model $f(x_t)$. We observe that the temporal evolution of the filtered estimates of the stochastic trend variable  closely follows  that of the observed prices. The filtered estimates for the storage model price component reveal that it predominantly captures periodically recurring price fluctuations with large price peaks and drops. Beyond the periods with elevated price volatility, the contribution of this component to the price variation appears small. This reflects that when equilibrium storage is an inner solution (so that $0<\sigma(x_t)<C$), the resulting price is subject to an intertemporal price restriction leading to prices which behave as a stationary Markov process.
Accordingly, in this no-arbitrage pricing regime, the economic storage model provides little additional information about the price evolution that goes beyond the stochastic trend.
However, storage becomes empirically relevant with a significant impact on the price behavior when the normal no-arbitrage pricing mechanism collapses in the stock-out and full-capacity regime, which occurs in the storage model in periods of severe and prolonged commodity shortages or oversupply.

The limits-to-arbitrage regimes (stock-out or full-capacity) detected by the fitted storage model  tend to coincide  with known historical market events.
For example, the  time periods with peaks in the filtered storage price component  for natural gas usually  correspond to periods when the historical level of natural gas storage in the market was very low \citep{kleppe_estimating_2017}.
The sharp drop in the storage price component for coffee in 1989 coincides with the collapse of the International Coffee Agreement (a cartel of coffee-producing countries) and oversupply in the market due to World Bank subsidies, while the 1994 peak is consistent with a negative supply shock triggered by significant frost damage in much of  the coffee-growing areas of Brazil.
The cotton price peak detected by the storage model in 2011 was arguably due to the severe global shortages, which were caused, inter alia, by the tightening of Indian export restrictions on cotton. The early nineties
spike in aluminum prices coincides with the collapse of the Soviet Union, and the 2008-2009 price drop is consistent  with the sharp decline in global aluminum demand that created a large stock overhang during this period after the subprime crisis.

\subsection{Model Comparisons}

In this section, we assess the empirical relevance of the price component related to the competitive storage model for explaining the observed price variation, and compare the storage SSM model with stochastic trend to  that with  deterministic  trend specifications. For this assessment, we rely on the marginal likelihood as well as diagnostic checks on Pearson and PIT residuals.

\subsubsection{Alternative Models}
For assessing the relevance of the storage model price component, we compare our SSM model to the restricted SSM that results in the absence of storage. The latter is obtained by letting $\delta \to 1$, making storage prohibitively costly, so that the stock process $x_t$ collapses to that of the supply shocks $z_t$. In this case the SSM in Equations (\ref{eq:Storagemodel})-(\ref{eq:Storagemodel-3}) with the assumed demand function $P(x)=\exp(-bx)$ reduces to
\begin{align*}
p_{t} & =k_{t}-bz_{t},\qquad z_{t}\sim\mbox{iid} N(0,1),\\
k_{t} & =k_{t-1}+\varepsilon_{t},\qquad\varepsilon_{t}\sim \mbox{iid} N(0,v^{2}).
\end{align*}
This represents a standard linear Gaussian local level (LGLL) SSM \citep{durbin_koopman_2ed} so that the Kalman filter can be applied for likelihood evaluation.
As the Kalman filter provides exact values for the likelihood, the PMMH used for simulating from the posterior of the parameters for the unrestricted storage SSM  can be replaced by a standard MH algorithm. The priors assigned to the two parameters $(b,v)$ are the same as those we assume for the unrestricted storage SSM.
\begin{table}
\centering 
 \resizebox{0.55\textwidth}{!}{ 
\begin{tabular}{lllll}
   \hline
  & Natgas & Coffee & Cotton & Aluminum \\ 
   \hline
Storage SSM & 164.15 & 420.07 & 522.80 & 545.36 \\ [0.3cm]
  LGLL SSM & 146.77 & 404.08 & 510.28 & 540.88 \\ 
    & (17.38) & (15.99) & (12.52) & (4.48) \\ [0.2cm]
  Linear trend & 109.57 & 309.64 & 427.00 & 496.77 \\ 
    & (54.58) & (110.43) & (95.80) & (48.59) \\ [0.2cm]
  RCS3 trend & 144.49 & 362.90 & 473.59 & 488.53 \\ 
    & (19.66) & (57.17) & (49.21) & (56.83) \\ [0.2cm]
  RCS7 trend & 132.03 & 375.50 & 488.44 & 517.23 \\ 
    & (32.12) & (44.57) & (34.36) & (28.13) \\ 
   \hline
\end{tabular}
 }

\caption{Log marginal likelihood values with the log Bayes factor of the storage SSM relative to the alternative models in parentheses.}
\label{tab:marglik}
\end{table}

As deterministic trend specifications to be compared with the stochastic trend in the storage SSM, we consider those used in the study of \cite{Gouel2017}. They use a linear time trend, for which $k_t$ in Equation (\ref{eq:Storagemodel-2}) is replaced by $k_t=\alpha+\beta t$. In addition, they consider restricted cubic spline trend specifications of the form $k_t=\sum_{g=1}^G=\gamma_g B_g(t)$, where $B_g(\cdot)$ are the basis functions of B-splines, $G$ is the degree of freedom, and $\gamma_g$ are the  corresponding trend  parameters to be estimated. For our comparison we consider restricted cubic splines with 3 knots (RSC3) and 5 trend parameters  as well as 7 knots (RSC7) and 9 trend parameters\footnote{The knots for the RSC3 specification are located at the 25\%, 50\% and 75\% quantiles of the time index and for the RSC7 at the 12.5\%, 25\%, 37.5\%, 50\%, 67.5\%, 75\% and 87.5\% quantiles.}. For these deterministic trends the SSM in Equations (\ref{eq:Storagemodel})-(\ref{eq:Storagemodel-3}) reduces to a univariate, non-linear autoregression for the log-price:
\begin{align}\label{eq:Determin-trend-models}
p_{t}=k_{t}+\log f\left[(1-\delta)\sigma(x_{t-1})+z_{t}\right],\qquad x_{t-1}=f^{-1}\left[\exp(p_{t-1}-k_{t-1})\right],\qquad z_t\sim \mbox{iid} N(0,1).
\end{align}
Analogously to the LGLL SSM, we can simulate from the posterior for the parameters of the deterministic trend models by using a standard MH algorithm.
For the structural parameters $(\delta,b)$  we assume the same priors as used in the storage SSM, and to the deterministic trend parameters $(\alpha,\beta,\gamma_g)$ we assign independent $N(0,20^2)$ priors. For details on the computation and derivation of the Pearson and PIT residuals of the deterministic trend models, see Appendix \ref{subsec:residuals}.

\subsubsection{Marginal Likelihood Model Comparisons and Diagnostics Checks}

\begin{figure}[h!]
\begin{centering}
\includegraphics[scale=0.4]{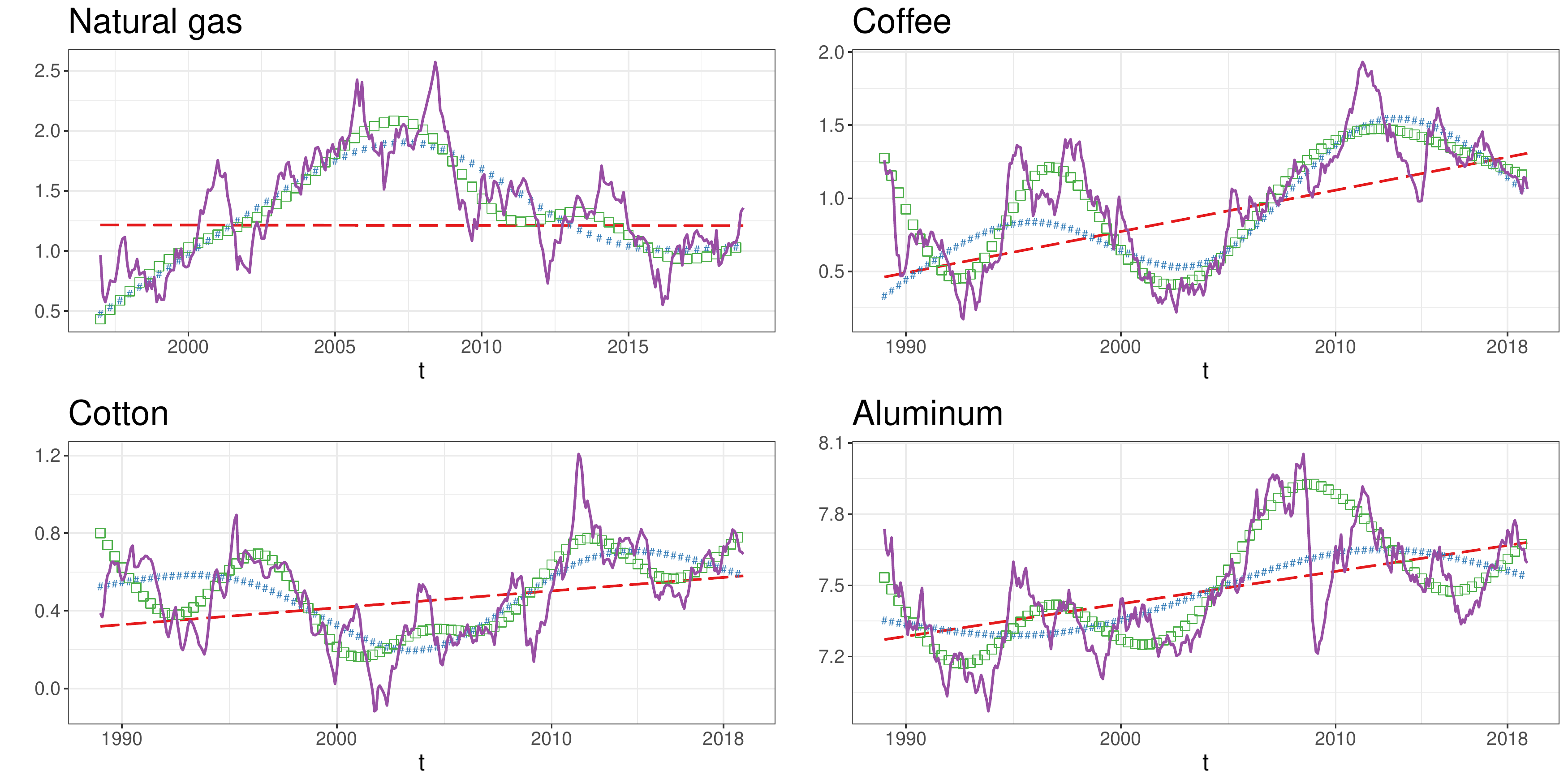}
\par\end{centering}
\caption{Fitted stochastic and deterministic trends. Smoothed stochastic trend (purple solid line), linear trend (red dashed line), RSC3 trend (blue number sign), RCS7 trend (green square).\label{fig:smoothed_trends}}
\end{figure}

Table \ref{tab:marglik} provides the log marginal likelihood values $\log \pi(p_{1:T}|\mbox{model}_s)$ for the storage SSM  together with those of the LGLL SSM and the storage model combined with the deterministic trend specifications. Also reported are the resulting values for the log Bayes factor of the storage SSM relative to the four alternative models $\log[\pi(p_{1:T}|\mbox{storage SSM})/\pi(p_{1:T}|\mbox{model}_\ell)]$.
The results reveal that the storage SSM is strongly preferred over the  LGLL SSM  for all commodities, which suggests that the structural storage component in the  SSM substantially contributes to the model fit. Hence, the non-linear price dynamics  with periodically recurring increases in price volatility and price spiking, as predicted by the competitive storage model, adds significantly to explaining the price behavior.
For all commodities, we also observe that the storage SSM is clearly favored over all deterministic trend specifications. Thus, the storage SSM has a trend component that is not only  consistent with the rationality of the economic model, but is also much more supported by the data than the deterministic trends, such as those used by \cite{Gouel2017} for the  estimation of  the structural parameters of the competitive storage model.
Our estimates of the structural parameters for the deterministic trend models  are found in Appendix  \ref{subsec:Additional_results}.  Figure \ref{fig:smoothed_trends} shows the time series plots of the fitted deterministic trends $\hat k_t$ and the smoothed mean of the stochastic trend $E(k_t|p_{1:T})$, all computed by setting the parameters to their  posterior mean values\footnote{The smoothed mean $E(k_t|p_{1:T})=p_t-E(\log f(x_t)|p_{1:T})$ is computed using the particle smoothing algorithm, which adds to the BPF as outlined in Section \ref{sec:BPF} a backward sampling step \citep[][Section 5]{doucet_tutorial_2009}.}. Unsurprisingly, we find that the stochastic trend  captures a substantially larger fraction of the observed price variations than the deterministic trends.

\begin{table}[t]
\begin{small}
\centering
\resizebox{0.7\textwidth}{!}{ 
\begin{tabular}{lcccccc}
 \hline\\[-0.2cm] 
  & Skew($\xi_t$)   & Kurt($\xi_t$)  & JB($\xi_t$) & $\rho_1(\eta_t)$ & LB$_{12}(\eta_t)$ & LB$_{12}(\eta_t^2)$ \\ \hline\\[-0.2cm] 
  & \multicolumn{6}{c}{Storage SSM} \\ 
 \cline{2-7}\\ 
 Natgas & 0.053 & 3.069 & 0.915 & 0.075 & 0.027 & 0.297 \\ 
  Coffee & 0.255 & 3.333 & 0.062 & 0.359 & $<$0.001 & $<$0.001 \\ 
  Cotton & -0.064 & 3.445 & 0.201 & 0.518 & $<$0.001 & $<$0.001 \\ 
  Aluminum & -0.214 & 3.25 & 0.159 & 0.291 & $<$0.001 & $<$0.001 \\ 
   [0.2cm]  & \multicolumn{6}{c}{LGLL SSM} \\ 
 \cline{2-7}\\ 
 Natgas & 0.033 & 4.298 & $<$0.001 & 0.084 & 0.055 & 0.452 \\ 
  Coffee & 0.801 & 7.681 & $<$0.001 & 0.258 & $<$0.001 & $<$0.001 \\ 
  Cotton & -0.23 & 6.325 & $<$0.001 & 0.502 & $<$0.001 & $<$0.001 \\ 
  Aluminum & -0.381 & 4.652 & $<$0.001 & 0.268 & $<$0.001 & $<$0.001 \\ 
   [0.2cm]  & \multicolumn{6}{c}{Linear trend} \\ 
 \cline{2-7}\\ 
 Natgas & 0.049 & 5.368 & $<$0.001 & 0.075 & 0.065 & 0.427 \\ 
  Coffee & -0.532 & 4.795 & $<$0.001 & 0.253 & $<$0.001 & 0.053 \\ 
  Cotton & 0.137 & 4.904 & $<$0.001 & 0.497 & $<$0.001 & $<$0.001 \\ 
  Aluminum & 0.625 & 6.502 & $<$0.001 & 0.243 & $<$0.001 & $<$0.001 \\ 
   [0.2cm]  & \multicolumn{6}{c}{RCS3 trend} \\ 
 \cline{2-7}\\ 
 Natgas & -0.099 & 3.962 & 0.005 & $<$0.001 & 0.003 & 0.068 \\ 
  Coffee & -0.508 & 5.231 & $<$0.001 & 0.185 & $<$0.001 & 0.004 \\ 
  Cotton & 0.094 & 4.669 & $<$0.001 & 0.449 & $<$0.001 & $<$0.001 \\ 
  Aluminum & 0.369 & 5.407 & $<$0.001 & 0.205 & $<$0.001 & 0.002 \\ 
   [0.2cm]  & \multicolumn{6}{c}{RCS7 trend} \\ 
 \cline{2-7}\\ 
 Natgas & -0.258 & 3.835 & 0.005 & 0.022 & $<$0.001 & 0.004 \\ 
  Coffee & -0.472 & 4.938 & $<$0.001 & 0.184 & $<$0.001 & 0.022 \\ 
  Cotton & 0.181 & 4.437 & $<$0.001 & 0.454 & $<$0.001 & $<$0.001 \\ 
  Aluminum & -0.059 & 3.495 & 0.144 & 0.198 & $<$0.001 & $<$0.001 \\ 
   [0.2cm] \hline 
 \end{tabular}
}
\caption{Diagnostics on the PIT and Pearson residuals. Skewness, Kurtosis, and $p$-value of the Jarque-Bera test (JB) for the PIT residuals. Lag-1 autocorrelation ($\rho_1$) and $p$-value of the Ljung-Box test (LB) for the Pearson residuals and their squared values, including 12 lags.}
\label{tab:diagnostics}
\end{small}
\end{table}

Table~\ref{tab:diagnostics} provides the results of diagnostic checks on the PIT residuals $\xi_t$ and the Pearson residuals $\eta_t$ for the storage SSM and the four alternative models considered. The PIT residuals of the storage SSM suggest that this model accounts well for the observed distributional properties of the prices for all commodities. The skewness and kurtosis of its PIT residuals  are close to their benchmark values for a normal distribution and they all pass the Jarque-Bera normality test at the 5\% significance level. In contrast, the LGLL SSM as well as the storage models with deterministic trends have difficulties approximating the distributional properties of the prices. Only the PIT residuals of the storage model with an RSC7 trend for aluminum  pass the Jarque-Bera normality test at a conventional significance level.

The first-order serial correlation of the Pearson residuals $\eta_t$ and the $p$-values of the Ljung-Box test for $\eta_t$ and $\eta_t^2$ including 12 lags reported in Table~\ref{tab:diagnostics} show that the storage SSM successfully accounts for the observed autocorrelation in the level and volatility of the gas price, while they point towards significant residual correlation in price level and volatility for coffee, cotton and aluminum.
However, all competing models cannot fully capture the serial correlation in the price levels of those three commodities either.
Only the volatility dynamics for coffee is better approximated by the linear and RSC7 trend model than by the storage SSM.
Clearly, based on these results, we can not identify  whether the failure of the storage SSM and the deterministic  trend models to explain all of the observed dynamics in the coffee, cotton and aluminum prices  is due to a potential misspecification of the trend or the competitive storage model itself, since the diagnostic tests are, as any specification test in this context, joint tests for the validity of both price components.

In sum, the results show that the storage SSM outperforms the deterministic trend models in explaining the observed distributional properties of  commodity prices, and  that its ability   to account for the dynamics in the price levels is not worse. Only in the approximation of the volatility dynamics, the deterministic trend specifications appear to have a slight advantage.

\subsubsection{Structural Parameter Estimates Under Stochastic and Deterministic Trends}

As it is evident from Figure 6, the dynamic and distributional characteristics of the de-trended prices substantially differ depending on whether a stochastic or deterministic trend is assumed. Therefore, it can be expected that the nature of the trend has a critical impact on the estimates of the parameters that determine the storage costs ($\delta$) and the price elasticity of demand ($b$), since these parameters are identified by the strength of the serial correlation and the size of the spikes in the trend-adjusted prices.
The lower the storage costs in the competitive storage model are, the stronger the predicted serial correlation, while the more inelastic the demand is, the larger the resulting price spikes.
As larger price spikes also imply more speculative storage activity, an inelastic demand  also contributes to the strength of the predicted serial correlation in the prices.

Table~\ref{tab:elacost} summarizes the estimates for the annualized storage costs (net of interest costs) in percent of the average price, and the price elasticities of demand obtained from the fitted storage SSM and the deterministic RCS trend models. The annual storage costs are computed as $-[(1-\delta)^{12}-1]$  and the price elasticity is given by $[-(b\bar x)^{-1}]$, where $\bar x$ is the mean supply.
We observe that the SSM with stochastic trend predicts substantially larger elasticities (in absolute values) than the deterministic trend models for all commodities and, except for natural gas, lower storage costs.
The larger elasticities found under the storage SSM reflect that the stochastic trend produces, due to its greater flexibility to track the observed price, trend-adjusted prices that have spikes that are smaller than those obtained under a deterministic trend.  Hence, in contrast to the deterministic trend specifications,  the stochastic trend SSM  is not forced to match the large spikes observed in the actual  prices by small estimated values for the elasticity.
For natural gas, the residual serial correlation in the prices adjusted by the stochastic trend component also appears to be relatively low, which indicates relatively high storage costs. However, for the other commodities, this residual serial correlation is larger  leading to substantially lower estimated storage costs.

\begin{table}
\begin{small}
\centering 
 \resizebox{0.85\textwidth}{!}{ 
 \begin{tabular}{lrrcrrcrrcrr}
 \hline 
 &\multicolumn{2}{c}{Natgas}  && \multicolumn{2}{c}{Coffee} && \multicolumn{2}{c}{Cotton} && \multicolumn{2}{c}{Aluminum}\\ 
 \cline{2-3} \cline{5-6} \cline{8-9} \cline{11-12} \\[-0.2cm] 
 &\multicolumn{1}{c}{costs} & \multicolumn{1}{c}{elast.} && \multicolumn{1}{c}{costs} & \multicolumn{1}{c}{elast.} && \multicolumn{1}{c}{costs} & \multicolumn{1}{c}{elast.} && \multicolumn{1}{c}{costs} & \multicolumn{1}{c}{elast.}\\ 
 \hline\\[-0.2cm] 
 Storage SSM & 12.6 & -1.03 &   & 2.7 & -0.65 &   & 1.6 & -0.69 &   & 1.3 & -1.46 \\ 
  RCS3 trend & 8.9 & -0.11 &   & 4.8 & -0.20 &   & 2.1 & -0.25 &   & 6.5 & -0.27 \\ 
  RCS7 trend & 11.1 & -0.10 &   & 4.1 & -0.19 &   & 4.7 & -0.26 &   & 1.7 & -0.30 \\ 
   \hline 
 \end{tabular} }

\caption{Estimates for the annual storage costs (net of interest costs) in percent of the average price and price elasticities of demand.}
\label{tab:elacost}
\end{small}
\end{table}

\cite{Gouel2017} provide estimates of storage costs and price elasticities of demand based on deterministic trend models used for annual data on various commodities, including coffee and cotton. This allows for some comparisons with our results for those two commodities.
The annual storage costs estimates they report for their preferred trend model for coffee and cotton are, respectively, 1.4\% and 0.3\% of the average price. These estimates based on annual data are much lower than those we found for the storage SSM as well as the deterministic trend models fitted to monthly data.
However, they argue that their estimated annual costs are possibly too small - an assessment that is consistent with our estimates for the  storage costs. For the annual price elasticity of demand, the estimates of \cite{Gouel2017} are -0.04\% for coffee and -0.03\% for cotton. These estimates imply a demand for those commodities which is substantially more inelastic than that implied from our estimates. One can argue which elasticities better reflect the markets. \cite{mehta2008responding} assume  a range of plausible values for the annual elasticity  of demand for coffee between  -0.2\% and -0.4\%, while \cite{duffy1990elasticity} argue that  the annual export demand for cotton is likely fairly elastic. Hence, our elasticity estimates are more in line with these assessments than those found by \cite{Gouel2017}.

\section{Conclusion}
In this paper, we have proposed a stochastic trend competitive storage model for commodity prices, which defines a non-linear state-space model (SSM). For the Bayesian posterior analysis of the proposed stochastic trend SSM, we use an efficient MCMC procedure. This adds to existing empirical commodity storage models based on deterministic trend specifications. Our stochastic trend approach fits into the economic rationality of the competitive storage model and is also sufficiently flexible to account for the variation in the observed prices that the competitive storage model is not intended to explain. The obvious benefit is that it makes the storage model applicable to markets with highly persistent unit root-like prices, which appears relevant for many commodity markets. Our approach aims at increasing the empirical relevance and applicability of the competitive storage model.

The MCMC procedure we propose for jointly estimating the structural and trend parameters in the SSM is a particle marginal Metropolis-Hastings algorithm based on the bootstrap particle filter. A Monte Carlo simulation experiment shows that this approach is able to disentangle the stochastic trend from the price variation due to speculative storage. The SSM is applied to monthly price data for natural gas, cotton, coffee and aluminum. Not surprisingly, the stochastic trend explains a large part of the observed variation in the commodity prices. More importantly, the competitive storage component adds short-run price volatility and price spiking, and becomes periodically relevant to explain non-linear pricing behavior related to states of market turmoil. A formal empirical comparison of the SSM to the corresponding model that results in the absence of storage suggests that the speculative storage price component significantly contributes to commodity price variation.

Which trend to apply will depend on the specific market under consideration.
If a stochastic trend is not appropriate, fitting a highly flexible stochastic trend model risks
overfitting the price variation and downplaying the contribution of
the storage model. Consequently, the price elasticity of demand will tend to be overestimated and the estimates of the storage costs can be expected to be correspondingly biased.
On the other hand, failing to account for a stochastic trend when it is appropriate  will tend
to underestimate the elasticity of demand. Our empirical results
show that the stochastic trend model consistently estimates a
higher elasticity of demand and a different amount of storage costs than existing deterministic
trend models for the commodity markets investigated in this paper.
Pre-testing of price characteristics can guide trend choice. For instance,
unit root tests can be applied to evaluate whether a stochastic trend
specification is suitable.

The empirical comparison of the stochastic trend SSM to existing deterministic trend models using the Bayes factor and model residual analysis shows that the stochastic trend fits the price data for the investigated commodity markets much better than the deterministic trends. In particular, in contrast to the deterministic trend specifications, the stochastic trend SSM captures the observed distributional properties of the prices, such as their skewness and kurtosis, quite well.
While the stochastic trend SSM also accounts for the price dynamics in the observed prices on the natural gas market it is not able to fully capture all the serial dependence of the coffee, cotton and aluminum prices.
This is similar to the results of financial approaches on modeling commodity term structures, showing the relevance of additional pricing factors beyond the traditional ones for the spot price and the convenience yield
(\citealt{miltersen1998pricing,schwartz1997stochastic,tang2012time}). The stochastic trend SSM is essentially a two-factor model with one reduced-form random walk component orthogonally appended to a factor restricted by economic constraints. Increasing the flexibility in the economic model will arguably improve the explanatory power of the model, although with additional statistical challenges in separately identifying the trend behavior from the price component related to the competitive storage model.
\bibliographystyle{chicago}
\bibliography{Storagebib}

\begin{thebibliography}{}

\bibitem[\protect\citeauthoryear{Andrieu, Doucet, and Holenstein}{Andrieu
  et~al.}{2010}]{andrieu_particle_2010}
Andrieu, C., A.~Doucet, and R.~Holenstein (2010).
\newblock Particle markov chain monte carlo methods.
\newblock {\em Journal of the Royal Statistical Society: Series B (Statistical
  Methodology)\/}~{\em 72\/}(3), 269--342.

\bibitem[\protect\citeauthoryear{Bobenrieth, Wright, and Zeng}{Bobenrieth
  et~al.}{2013}]{bobenrieth2013stocks}
Bobenrieth, E., B.~Wright, and D.~Zeng (2013).
\newblock Stocks-to-use ratios and prices as indicators of vulnerability to
  spikes in global cereal markets.
\newblock {\em Agricultural Economics\/}~{\em 44\/}(s1), 43--52.

\bibitem[\protect\citeauthoryear{Bos and Shephard}{Bos and
  Shephard}{2006}]{BosShephard2006}
Bos, C.~S. and N.~Shephard (2006).
\newblock Inference for adaptive time series models: Stochastic volatility and
  conditionally gaussian state space form.
\newblock {\em Econometric Reviews\/}~{\em 25\/}(2-3), 219--244.

\bibitem[\protect\citeauthoryear{Cafiero, Bobenrieth~H., and
  Bobenrieth~H.}{Cafiero et~al.}{2011}]{cafiero2011storage}
Cafiero, C., E.~Bobenrieth~H., and J.~Bobenrieth~H. (2011).
\newblock Storage arbitrage and commodity price volatility.
\newblock {\em Safeguarding food security in volatile global markets\/},
  301--326.

\bibitem[\protect\citeauthoryear{Cafiero, Bobenrieth~H., Bobenrieth~H., and
  Wright}{Cafiero et~al.}{2011}]{cafiero2011empirical}
Cafiero, C., E.~Bobenrieth~H., J.~Bobenrieth~H., and B.~D. Wright (2011).
\newblock {The empirical relevance of the competitive storage model}.
\newblock {\em Journal of Econometrics\/}~{\em 162\/}(1), 44--54.

\bibitem[\protect\citeauthoryear{Cafiero, Bobenrieth~H., Bobenrieth~H., and
  Wright}{Cafiero et~al.}{2015}]{cafiero2015maximum}
Cafiero, C., E.~Bobenrieth~H., J.~Bobenrieth~H., and B.~D. Wright (2015).
\newblock Maximum likelihood estimation of the standard commodity storage
  model: Evidence from sugar prices.
\newblock {\em American Journal of Agricultural Economics\/}~{\em 97\/}(1),
  122--136.

\bibitem[\protect\citeauthoryear{Canova}{Canova}{2014}]{canova2014bridging}
Canova, F. (2014).
\newblock Bridging dsge models and the raw data.
\newblock {\em Journal of Monetary Economics\/}~{\em 67}, 1--15.

\bibitem[\protect\citeauthoryear{Capp\'e, Godsill, and Moulines}{Capp\'e
  et~al.}{2007}]{Cappe2007}
Capp\'e, O., S.~J. Godsill, and E.~Moulines (2007).
\newblock An overview of existing methods and recent advances in sequential
  monte carlo.
\newblock {\em Proceedings of the IEEE\/}~{\em 95\/}(5), 899--924.

\bibitem[\protect\citeauthoryear{Chib and Jeliazkov}{Chib and
  Jeliazkov}{2001}]{ChibJeliazkov2001}
Chib, S. and I.~Jeliazkov (2001).
\newblock {Marginal likelihood from the Metropolis-Hastings output}.
\newblock {\em Journal of the American Statistical Association\/}~{\em
  96\/}(453), 270--281.

\bibitem[\protect\citeauthoryear{Deaton and Laroque}{Deaton and
  Laroque}{1992}]{deaton_behaviour_1992}
Deaton, A. and G.~Laroque (1992).
\newblock On the behaviour of commodity prices.
\newblock {\em The Review of Economic Studies\/}~{\em 59\/}(1), 1--23.

\bibitem[\protect\citeauthoryear{Deaton and Laroque}{Deaton and
  Laroque}{1995}]{Deaton1995}
Deaton, A. and G.~Laroque (1995).
\newblock Estimating a nonlinear rational expectations commodity price model
  with unobservable state variables.
\newblock {\em Journal of Applied Econometrics\/}~{\em 10\/}(S1), S9--S40.

\bibitem[\protect\citeauthoryear{Deaton and Laroque}{Deaton and
  Laroque}{1996}]{Deaton1996}
Deaton, A. and G.~Laroque (1996).
\newblock {Competitive Storage and Commodity Price Dynamics}.
\newblock {\em The Journal of Political Economy\/}~{\em 104(5)}, 896--923.

\bibitem[\protect\citeauthoryear{DeJong and Dave}{DeJong and
  Dave}{2011}]{deJong2007}
DeJong, D.~N. and C.~Dave (2011).
\newblock {\em Structural Macroeconometrics\/} (2 ed.).
\newblock Princeton University Press.

\bibitem[\protect\citeauthoryear{Doucet and Johansen}{Doucet and
  Johansen}{2009}]{doucet_tutorial_2009}
Doucet, A. and A.~M. Johansen (2009).
\newblock A tutorial on particle filtering and smoothing: {Fifteen} years
  later.
\newblock {\em Handbook of nonlinear filtering\/}~{\em 12\/}(656-704), 3.

\bibitem[\protect\citeauthoryear{Duffy, Wohlgenant, and Richardson}{Duffy
  et~al.}{1990}]{duffy1990elasticity}
Duffy, P.~A., M.~K. Wohlgenant, and J.~W. Richardson (1990).
\newblock The elasticity of export demand for us cotton.
\newblock {\em American Journal of Agricultural Economics\/}~{\em 72\/}(2),
  468--474.

\bibitem[\protect\citeauthoryear{Durbin and Koopman}{Durbin and
  Koopman}{2012}]{durbin_koopman_2ed}
Durbin, J. and S.~J. Koopman (2012).
\newblock {\em Time Series Analysis by State Space Methods\/} (2 ed.).
\newblock Number~38 in Oxford Statistical Science. Oxford University Press.

\bibitem[\protect\citeauthoryear{Flury and Shephard}{Flury and
  Shephard}{2011}]{flury_shephart_2011}
Flury, T. and N.~Shephard (2011).
\newblock Bayesian inference based only on simulated likelihood: Particle
  filter analysis of dynamic economic models.
\newblock {\em Econometric Theory\/}~{\em 27\/}(Special Issue 05), 933--956.

\bibitem[\protect\citeauthoryear{Geyer}{Geyer}{1992}]{geyer1992}
Geyer, C.~J. (1992).
\newblock Practical {M}arkov chain {M}onte {C}arlo.
\newblock {\em Statistical Science\/}~{\em 7\/}(4), 473--483.

\bibitem[\protect\citeauthoryear{Gordon, Salmond, and Smith}{Gordon
  et~al.}{1993}]{Gordon1993}
Gordon, N.~J., D.~J. Salmond, and A.~F.~M. Smith (1993).
\newblock Novel approach to nonlinear/non-gaussian bayesian state estimation.
\newblock {\em IEE Proceedings F - Radar and Signal Processing\/}~{\em
  140\/}(2), 107--113.

\bibitem[\protect\citeauthoryear{Gouel and Legrand}{Gouel and
  Legrand}{2017}]{Gouel2017}
Gouel, C. and N.~Legrand (2017).
\newblock Estimating the competitive storage model with trending commodity
  prices.
\newblock {\em Journal of Applied Econometrics\/}~{\em 32\/}(4), 744--763.

\bibitem[\protect\citeauthoryear{Guerra, Bobenrieth~H., Bobenrieth~H., and
  Cafiero}{Guerra et~al.}{2015}]{guerra2015empirical}
Guerra, V., E.~Bobenrieth~H., J.~Bobenrieth~H., and C.~Cafiero (2015).
\newblock Empirical commodity storage model: the challenge of matching data and
  theory.
\newblock {\em European Review of Agricultural Economics\/}~{\em 42\/}(4),
  607--623.

\bibitem[\protect\citeauthoryear{Gustafson}{Gustafson}{1958}]{gustafson1958carryover}
Gustafson, R.~L. (1958).
\newblock {\em Carryover levels for grains: a method for determining amounts
  that are optimal under specified conditions}.
\newblock Number 1178. US Department of Agriculture.

\bibitem[\protect\citeauthoryear{Haario, Saksman, and Tamminen}{Haario
  et~al.}{2001}]{haario_adaptive_2001}
Haario, H., E.~Saksman, and J.~Tamminen (2001).
\newblock An adaptive {Metropolis} algorithm.
\newblock {\em Bernoulli\/}~{\em 7\/}(2), 223--242.

\bibitem[\protect\citeauthoryear{Kim, Shephard, and Chib}{Kim
  et~al.}{1998}]{KimShephardChib1998}
Kim, S., N.~Shephard, and S.~Chib (1998).
\newblock Stochastic volatility: Likelihood inference and comparison with arch
  models.
\newblock {\em The Review of Economic Studies\/}~{\em 65\/}(3), 361--393.

\bibitem[\protect\citeauthoryear{Kleppe and Oglend}{Kleppe and
  Oglend}{2017}]{kleppe_estimating_2017}
Kleppe, T.~S. and A.~Oglend (2017).
\newblock Estimating the competitive storage model: {A} simulated likelihood
  approach.
\newblock {\em Econometrics and Statistics\/}~{\em 4}, 39--56.

\bibitem[\protect\citeauthoryear{Kleppe and Oglend}{Kleppe and
  Oglend}{2019}]{oglend2019}
Kleppe, T.~S. and A.~Oglend (2019).
\newblock Can limits-to-arbitrage from bounded storage improve commodity
  term-structure modeling?
\newblock {\em Journal of Futures Markets\/}~{\em 39\/}(7), 865--889.

\bibitem[\protect\citeauthoryear{Legrand}{Legrand}{2019}]{legrandempirical}
Legrand, N. (2019).
\newblock The empirical merit of structural explanations of commodity price
  volatility: Review and perspectives.
\newblock {\em Journal of Economic Surveys\/}~{\em 33\/}(2), 639--664.

\bibitem[\protect\citeauthoryear{Liesenfeld and Richard}{Liesenfeld and
  Richard}{2003}]{liesenfeld_richard_03}
Liesenfeld, R. and J.-F. Richard (2003).
\newblock Univariate and multivariate stochastic volatility models: estimation
  and diagnostics.
\newblock {\em Journal of Empirical Finance\/}~{\em 10\/}(4), 505--531.

\bibitem[\protect\citeauthoryear{Mehta and Chavas}{Mehta and
  Chavas}{2008}]{mehta2008responding}
Mehta, A. and J.-P. Chavas (2008).
\newblock Responding to the coffee crisis: What can we learn from price
  dynamics?
\newblock {\em Journal of Development Economics\/}~{\em 85\/}(1-2), 282--311.

\bibitem[\protect\citeauthoryear{Miltersen and Schwartz}{Miltersen and
  Schwartz}{1998}]{miltersen1998pricing}
Miltersen, K.~R. and E.~S. Schwartz (1998).
\newblock {Pricing of options on commodity futures with stochastic term
  structures of convenience yields and interest rates}.
\newblock {\em Journal of Financial and Quantitative Analysis\/}~{\em 33\/}(1),
  33--59.

\bibitem[\protect\citeauthoryear{Oglend and Kleppe}{Oglend and
  Kleppe}{2017}]{oglend_behavior_2017}
Oglend, A. and T.~S. Kleppe (2017).
\newblock On the behavior of commodity prices when speculative storage is
  bounded.
\newblock {\em Journal of Economic Dynamics and Control\/}~{\em 75}, 52--69.

\bibitem[\protect\citeauthoryear{Richard and Zhang}{Richard and
  Zhang}{2007}]{richardetal07}
Richard, J.-F. and W.~Zhang (2007).
\newblock Efficient high-dimensional importance sampling.
\newblock {\em Journal of Econometrics\/}~{\em 127\/}(2), 1385--1411.

\bibitem[\protect\citeauthoryear{Routledge, Seppi, and Spatt}{Routledge
  et~al.}{2000}]{Routledge2000}
Routledge, B.~R., D.~J. Seppi, and C.~S. Spatt (2000).
\newblock Equilibrium forward curves for commodities.
\newblock {\em The Journal of Finance\/}~{\em 55\/}(3), 1297--1337.

\bibitem[\protect\citeauthoryear{Sala}{Sala}{2015}]{sala2015dsge}
Sala, L. (2015).
\newblock Dsge models in the frequency domains.
\newblock {\em Journal of Applied Econometrics\/}~{\em 30\/}(2), 219--240.

\bibitem[\protect\citeauthoryear{Schwartz}{Schwartz}{1997}]{schwartz1997stochastic}
Schwartz, E.~S. (1997).
\newblock {The stochastic behavior of commodity prices: Implications for
  valuation and hedging}.
\newblock {\em The Journal of Finance\/}~{\em 52\/}(3), 923--973.

\bibitem[\protect\citeauthoryear{Shephard and Pitt}{Shephard and
  Pitt}{1997}]{shepard_pitt97}
Shephard, N. and M.~K. Pitt (1997).
\newblock Likelihood analysis of non-{G}aussian measurement time series.
\newblock {\em Biometrika\/}~{\em 84}, 653--667.

\bibitem[\protect\citeauthoryear{Tang}{Tang}{2012}]{tang2012time}
Tang, K. (2012).
\newblock {Time-varying long-run mean of commodity prices and the modeling of
  futures term structures}.
\newblock {\em Quantitative Finance\/}~{\em 12\/}(5), 781--790.

\bibitem[\protect\citeauthoryear{Wang and Tomek}{Wang and
  Tomek}{2007}]{wang2007commodity}
Wang, D. and W.~G. Tomek (2007).
\newblock {Commodity prices and unit root tests}.
\newblock {\em American Journal of Agricultural Economics\/}~{\em 89\/}(4),
  873--889.

\end{thebibliography}

\newpage
\appendix

\renewcommand{\thefigure}{A-\arabic{figure}} 
\setcounter{figure}{0}  
\renewcommand{\theequation}{A-\arabic{equation}} 
\setcounter{equation}{0}  
\renewcommand{\thetable}{A-\arabic{table}} 
\setcounter{table}{0}  

\newpage

\section{Appendix\label{sec:Appendix}}

\subsection{Numerical Solution of the Price Function\label{subsec:Price-function}}
The numerical algorithm we use to solve  the functional equation for the price function $f(x)$ as defined by Equations (\ref{eq:fx})-(\ref{eq:sig_f_rel}) is based on that used by \citet{oglend2019} for a model with autocorrelated supply shocks. The algorithm is based on solving the storage policy function  $\sigma(x)$ and then recover $f(x)$ via Equation (\ref{eq:sig_f_rel}), which implies that
\begin{equation}
f(x)=P(x-\sigma(x)).\label{eq:pricefunc}
\end{equation}

Let $x^{*}$ be defined such that $P(x^{*})=\bar{f}(x^{*})$ and
$x^{**}$  such that $P(x^{**}-C)=\bar{f}(x^{**})$. For $x<x^{*}$, so that
$f(x)=P(x)$, it follows that $\sigma(x)=0$ (stock-out regime); for $x^{*}\leq x\leq x^{**}$,
so that $f(x)=\bar{f}(x)$, it follows that $\sigma(x)\in[0,C]$ (storage regime);
and for $x>x^{**}$, so that $f(x)=P(x-C)$, it follows that  $\sigma(x)=C$ (full
capacity storage regime).

The numerical representation of $\sigma(x)$ is given by $\mathcal{S}=\{\hat{x}^{*},\hat{x}^{**},s(x)\}$,
where the function $s(x)$ (with $s(x)\simeq\sigma(x)$ for $x\in[\hat{x}^{*},\hat{x}^{**}]$)
is represented on a (comparatively sparse) grid on $[\hat{x}^{*},\hat{x}^{**}]$
and is evaluated using a suitable  interpolation method (e.g. linear).
The resulting approximation is given by
\[
\sigma_{\mathcal{S}}(x)=\begin{cases}
0 & \text{ if }x<\hat{x}^{*}\\
s(x) & \text{ if }\hat{x}^{*}\leq x\leq\hat{x}^{**}\\
C & \text{ if }x>\hat{x}^{**}
\end{cases},
\]
and correspondingly $f_{\mathcal{S}}(x)=P(x-\sigma_{\mathcal{S}}(x))$.

The iteration to find $\hat{\sigma}_{\mathcal{S}}(x)\simeq\sigma(x)$
consists of the following steps:
\begin{enumerate}
\item Select an initial guess, e.g. $\mathcal{S}_{1}=\{\hat{x}_{1}^{*},\hat{x}_{1}^{**},s_{1}(x)\}=\{0,C,s_{1}(x)\}$,
where $s_{1}(x)$ is the linear function such that $s(0)=0$, $s(C)=C$.
Set $n=1$.
\item Update the left kink point $\hat{x}_{n+1}^{*}$ according to
\begin{equation}
\hat{x}_{n+1}^{*}=D\left(\beta\int f_{\mathcal{S}_{n}}(z)\phi(z)dz\right).\label{eq:nsolver_1}
\end{equation}
\item Update the right kink point $\hat{x}_{n+1}^{**}$ according to
\begin{equation}
\hat{x}_{n+1}^{**}=D\left(\beta\int f_{\mathcal{S}_{n}}((1-\delta)C+z)\phi(z)dz\right)+C.\label{eq:nsolver_2}
\end{equation}
\item Update the grid $\{x_{n+1}^{(j)}\}$ to be on $[\hat{x}_{n+1}^{*},\hat{x}_{n+1}^{**}].$
\item For each grid point $j$, find the update $s_{n+1}(x_{n+1}^{(j)})$ as
the solution in $s$ to
\begin{equation}
s=x_{n+1}^{(j)}-D\left(\beta\int f_{\mathcal{S}_{n}}((1-\delta)s+z)\phi(z)dz\right),\label{eq:nsolver_3}
\end{equation}
using a univariate non-linear root-finding algorithm. Notice that the solution
$s=s_{n+1}(x_{n+1}^{(j)})$ is constrained to be in $[0,C]$, and
that $s_{n+1}(\hat{x}_{n+1}^{*})=0$, $s_{n+1}(\hat{x}_{n+1}^{**})=C$.
\item Until convergence, set $n\leftarrow n+1$ and go back to step 2.
\end{enumerate}
The integrals in Equations (\ref{eq:nsolver_1})-(\ref{eq:nsolver_3})
are approximated using the trapezoidal quadrature rule with 128 subintervals,
over the interval $[-4,4]$, and the non-linear equation~(\ref{eq:nsolver_3})
is solved using Brent's method. Allowing the grid space to adjust
to the updated functional solutions ensures that the grid can  dynamically
concentrate in the region of the state-space where a high precision
is needed, namely the region defining the storage regime. This provides both efficient
and precise numerical solutions to the pricing function.

\subsection{Residuals for the Deterministic Trend Models\label{subsec:residuals}}
For the deterministic trend models as given by Equation (\ref{eq:Determin-trend-models}) the conditional expectation $E(p_{t+1}|p_{1:t})$ and variance $\text{Var}(p_{t+1}|p_{1:t})$ defining the Pearson residuals in Equation (\ref{eq:Pearson}) can be evaluated by MC integration as the sample mean and variance of the simulated prices
\begin{align}
p_{t+1}^k =k_{t+1}+\log f[(1-\delta)\sigma(x_t)+z_{t+1}^k], \qquad k=1,\ldots,N,
\end{align}
where $\{z_{t+1}^k\}_{k=1}^N$ are iid draws from a $N(0,1)$ distribution.

The PIT residuals in Equation (\ref{eq:PIT}) obtain as follows: The probability $u_{t+1}=\mbox{Pr}(p_{t+1}\leq p_{t+1}^{o}|p_{1:t})$, which follows for a correctly specified model a uniform distribution $U_{[0,1]}$ on the unit interval,  results as
\begin{align}
u_{t+1}&=\mbox{Pr}(\exp(p_{t+1}-k_{t+1} )\leq \exp(p_{t+1}^{o}-k_{t+1})\,|\,p_{1:t})\\
       &=\mbox{Pr}( f^{-1}[\exp(p_{t+1}-k_{t+1} )]\geq f^{-1}[\exp(p_{t+1}^{o}-k_{t+1})]\,|\,p_{1:t})\label{eq:app_pit1}\\
        &=\mbox{Pr}( z_{t+1}\geq z_{t+1}^{o} \,|\,p_{1:t}), \label{eq:app_pit2}
\end{align}
where $z_{t+1}^{o}=f^{-1}[\exp(p_{t+1}^{o}-k_{t+1})] -(1-\delta)\sigma(x_t)$. Equation (\ref{eq:app_pit1}) follows from the fact that the inverse of the rational expectations equilibrium price function $f^{-1}$ is monotonically non-increasing. Since $z_t$ is a $N(0,1)$ random variate with cdf denoted by $\Phi$, Equation (\ref{eq:app_pit2}) implies that $1-u_{t+1}=\Phi(z_{t+1}^{(o)})$. Since for $u_{t+1}\sim U_{[0,1]}$ it holds that  $1-u_{t+1}\sim U_{[0,1]}$, the PIT residuals are given by
\begin{align}
\xi_{t+1}&=\Phi^{-1}(1-u_{t+1})\\
         &=\Phi^{-1}(\Phi(z_{t+1}^{(o)}))\\
         &=z_{t+1}^{(o)}.
\end{align}

\subsection{Additional Estimation Results \label{subsec:Additional_results}}
Table \ref{tab:pars_all} provides the posterior mean and standard deviation  for the structural parameters of the competitive storage model combined  with deterministic trend specifications.
\begin{table}[H]
\centering 
 \resizebox{0.65\textwidth}{!}{ 
\begin{tabular}{lllllll}
   \hline
 &  &  & Natgas & Coffee & Cotton & Aluminum \\ 
   \hline
$\delta$ & Linear & Post. mean & 0.0090 & 0.0025 & 0.0023 & 0.0087 \\ 
   &  & Post. std. & 0.0053 & 0.0016 & 0.0015 & 0.0030 \\ 
   &  &   &  &  &  &  \\ 
   & RCS3 & Post. mean & 0.0077 & 0.0041 & 0.0017 & 0.0056 \\ 
   &  & Post. std. & 0.0044 & 0.0023 & 0.0012 & 0.0021 \\ 
   &  &   &  &  &  &  \\ 
   & RCS7 & Post. mean & 0.0098 & 0.0035 & 0.0040 & 0.0014 \\ 
   &  & Post. std. & 0.0051 & 0.0024 & 0.0022 & 0.0010 \\ 
   &  &   &  &  &  &  \\ 
  $b$ & Linear & Post. mean & 2.05 & 1.33 & 1.02 & 0.75 \\ 
   &  & Post. std. & 0.122 & 0.058 & 0.045 & 0.036 \\ 
   &  &   &  &  &  &  \\ 
   & RCS3 & Post. mean & 1.83 & 1.12 & 0.84 & 0.74 \\ 
   &  & Post. std. & 0.110 & 0.056 & 0.045 & 0.046 \\ 
   &  &   &  &  &  &  \\ 
   & RCS7 & Post. mean & 1.82 & 1.08 & 0.77 & 0.67 \\ 
   &  & Post. std. & 0.118 & 0.054 & 0.043 & 0.032 \\ 
   \hline
\end{tabular}
 }

\caption{Estimates for the  storage model parameters
under the storage model with deterministic trends. }\label{tab:pars_all}
\end{table}

\end{document}